\documentclass[10pt, prd, twocolumn, nofootinbib,preprint,superscriptaddress]{revtex4}
\pdfoutput=1
\usepackage[T1]{fontenc}
\usepackage{amsmath,amssymb}
\usepackage{epsfig}
\usepackage{graphicx}
\usepackage[usenames,dvipsnames]{color}
\usepackage{subfigure}
\usepackage{slashed}
\usepackage[colorlinks,citecolor=blue]{hyperref}
\usepackage{pdfpages}
\usepackage{color}

\begin{document}
	\title{Muon $(g-2)$ and XENON1T Excess with Boosted Dark Matter in $L_{\mu}-L_{\tau}$ Model}

	\author{Debasish Borah}
	\email{dborah@iitg.ac.in}
	\affiliation{Department of Physics, Indian Institute of Technology Guwahati, Assam 781039, India}
	
	\author{Manoranjan Dutta}
	\email{ph18resch11007@iith.ac.in}
	\affiliation{Department of Physics, Indian Institute of Technology Hyderabad, Kandi, Sangareddy 502285, Telangana, India}
	
	\author{Satyabrata Mahapatra}
	\email{ph18resch11001@iith.ac.in}
	\affiliation{Department of Physics, Indian Institute of Technology Hyderabad, Kandi, Sangareddy 502285, Telangana, India}
	
	\author{Narendra Sahu}
	\email{nsahu@iith.ac.in}
	\affiliation{Department of Physics, Indian Institute of Technology Hyderabad, Kandi, Sangareddy 502285, Telangana, India}

	\begin{abstract}
		Motivated by the growing evidence for lepton flavour universality violation after the first results from Fermilab's muon $(g-2)$ measurement, we revisit one of the most widely studied anomaly free extensions of the standard model namely, gauged $L_{\mu}-L_{\tau}$ model, known to be providing a natural explanation for muon $(g-2)$. We also incorporate the presence of dark matter (DM) in this model in order to explain the recently reported electron recoil excess by the XENON1T collaboration. We show that the same neutral gauge boson responsible for generating the required muon $(g-2)$ can also mediate interactions between electron and dark fermions boosted by dark matter annihilation. The required DM annihilation rate into dark fermion require a hybrid setup of thermal and non-thermal mechanisms to generate DM relic density. The tightly constrained parameter space from all requirements remain sensitive to ongoing and near future experiments, keeping the scenario very predictive.
	\end{abstract}
	
	\maketitle
	
	\noindent
	{\bf Introduction}: The recent measurement of the muon anomalous magnetic moment, $a_\mu$ = $(g - 2)_\mu/2$, by the E989
	experiment at Fermilab shows a discrepancy with respect to the theoretical prediction of the Standard
	Model (SM) \cite{Abi:2021gix}
	\begin{eqnarray}
		a^{\rm FNAL}_\mu = 116 592 040(54) \times 10^{-11}\\
		a^{\rm SM}_\mu = 116 591 810(43) \times 10^{-11}
	\end{eqnarray}
	which when combined with the previous Brookhaven determination of
	\begin{equation}
		a^{\rm BNL}_\mu = 116 592 089(63) \times 10^{-11}
	\end{equation}
	leads to a 4.2 $\sigma$ observed excess of
	$\Delta a_\mu = 251(59) \times 10^{-11}$.
	The status of the SM calculation of muon magnetic moment has been updated recently in \cite{Aoyama:2020ynm}\footnote{The latest lattice results \cite{Borsanyi:2020mff} however, predict a larger value of muon $(g-2)$ bringing it closer to experimental value.}. For more details, one may refer to \cite{Zyla:2020zbs, Lindner:2016bgg,Davier:2019can,Davier:2017zfy,Davier:2010nc}. The latest Fermilab measurements have also led to several recent works on updating possible theoretical models with new data. For example, see \cite{Arcadi:2021cwg,Zhu:2021vlz,Han:2021gfu, Baum:2021qzx, Bai:2021bau, Das:2021zea, Lu:2021vcp} for minimal dark matter (DM) motivated scenarios, \cite{Ge:2021cjz, Brdar:2021pla, Buen-Abad:2021fwq} for axion like particle (ALP) motivated scenarios, \cite{Zu:2021odn, Amaral:2021rzw} for gauged lepton flavour models like $U(1)_{L_{\mu}-L_{\tau}}$ and \cite{Endo:2021zal, Ahmed:2021htr, Abdughani:2021pdc, VanBeekveld:2021tgn, Cox:2021gqq, Wang:2021bcx, Gu:2021mjd, Cao:2021tuh, Yin:2021mls, Han:2021ify, Aboubrahim:2021rwz, Yang:2021duj,Chakraborti:2021bmv, Ferreira:2021gke, Wang:2021fkn, Li:2021poy, Cadeddu:2021dqx, Calibbi:2021qto, Chen:2021vzk, Escribano:2021css, Chun:2021dwx, Arcadi:2021yyr, Chen:2021jok, Nomura:2021oeu} for other phenomenological scenarios like supersymmetry, multi-Higgs doublet models etc. and other implications of this new measurement. For a comprehensive review on new physics explanations of muon $(g-2)$ anomaly, please see \cite{Athron:2021iuf}. Another evidence of such lepton flavour universality (LFU) violation, that too in the context of muon, comes from the measurement of $R_K = {\rm BR}(B \rightarrow K \mu^+ \mu^-)/{\rm BR}(B \rightarrow K e^+ e^-)$. While the hint for this anomaly, like muon $(g-2)$ was there for several years, recent update from the LHCb collaboration \cite{Aaij:2021vac} has led to the most precise measurement ever with more than $3\sigma$ deviation from the SM predictions. In the light of growing evidences for such LFU violations, need for beyond standard model physics around the TeV corner has become very prominent.
	
	Another recent anomaly is the one reported by the XENON1T collaboration in 2020 related to their observation of an excess of electron recoil events over the background in the recoil energy $E_r$ in a range 1-7 keV,	peaked around 2.4 keV\cite{Aprile:2020tmw}. Although solar axions and neutrinos with magnetic moment can explain the excess at $3.5\sigma$ and $3.2\sigma$ significance respectively, they are severely plagued by stellar cooling bounds. This has led to several interesting new physics explanations, see \cite{Takahashi:2020bpq,Alonso-Alvarez:2020cdv,Kannike:2020agf,Fornal:2020npv,Du:2020ybt,Ko:2020gdg, Su:2020zny,Harigaya:2020ckz, Borah:2020jzi,Choudhury:2020xui, Bramante:2020zos, Bell:2020bes,  Borah:2020smw, Aboubrahim:2020iwb, Lee:2020wmh, Baek:2020owl, Shakeri_2020, Bally:2020yid, DelleRose:2020pbh, Ema:2020fit, Dutta:2021wbn} and references therein. The DM interpretations out of these examples, typically have a light mediator via which DM interacts with electrons. The recoil can occur either due to light boosted DM or inelastic up or down-scattering~\cite{Bell:2020bes, Lee:2020wmh, Baek:2020owl, Harigaya:2020ckz, Bramante:2020zos, Baryakhtar:2020rwy, Chao:2020yro, An:2020tcg, He:2020wjs, Choudhury:2020xui, Borah:2020jzi, Kim:2020aua, Shakeri_2020, Borah:2020smw, Keung:2020uew, Aboubrahim:2020iwb, He:2020sat, Choi:2020ysq, McKeen:2020vpf, Jho:2020sku, Alhazmi:2020fju, Jho:2020sku, Dutta:2021wbn, Das:2021lcr}.
	
	Here we consider the popular and minimal model based on the gauged $L_{\mu}-L_{\tau}$ symmetry which is anomaly free \cite{He:1990pn, He:1991qd}. In earlier attempts to explain XENON1T excess with inelastic DM in gauged $L_{\mu}-L_{\tau}$ model \cite{Borah:2020jzi} which can also explain $(g-2)_{\mu}$, only a tiny parameter space was allowed from all requirements even while considering a much larger error bars in $(g-2)_{\mu}$ namely $\Delta a_{\mu}  = (27.9 \pm 22.8)\times 10^{-10}$, consistent with the 3.7$\sigma$ discrepancy prior to the Fermilab measurement. As can be seen from \cite{Borah:2020jzi}, the main obstacle in satisfying both the excess is the constraint on heavier DM lifetime. To be more specific, in such scenarios, the heavier DM must be present in the universe at current epoch so that it can give rise to inelastic down-scattering at XENON1T detector. However, the same process responsible for such scattering also leads to heavier DM decay into lighter DM and SM particles leading to stringent constraints. Therefore, in this work, we consider a single component DM scenario which can annihilate into boosted lighter particles so that the latter can scatter off electron elastically, giving rise to the required excess. Boosted DM interpretation of XENON1T excess in the context of different models have been discussed in \cite{Kannike:2020agf,Fornal:2020npv,Du:2020ybt,Ko:2020gdg, McKeen:2020vpf, Jho:2020sku, Alhazmi:2020fju, Jho:2020sku, Das:2021lcr} \footnote{See \cite{Kim:2016zjx, Giudice:2017zke} for earlier works on this possibility.}. We study this possibility within the framework of gauged $L_{\mu}-L_{\tau}$ model along with the possibility of explaining the muon $(g-2)$ data. While we do not pursue the study of $R_K$ anomalies in this model, one may refer to \cite{Biswas:2019twf} for common origin of muon $(g-2)$ and $R_K$ anomalies along with dark matter in extensions of minimal $L_{\mu}-L_{\tau}$ model.\\

	\noindent
	{\bf Gauged $L_{\mu}-L_{\tau}$ Symmetry}:
	The SM fermion content with their gauge charges under $SU(3)_c \times SU(2)_L \times U(1)_Y \times U(1)_{L_{\mu}-L_{\tau}}$ gauge symmetry are denoted as follows.
	
	$$ q_L=\begin{pmatrix}u_{L}\\
		d_{L}\end{pmatrix} \sim (3, 2, \frac{1}{6}, 0), \; u_R (d_R) \sim (3, 1, \frac{2}{3} (-\frac{1}{3}), 0)$$
	$$L_e=\begin{pmatrix}\nu_{e}\\
		e_{L}\end{pmatrix} \sim (1, 2, -\frac{1}{2}, 0), \; e_R \sim (1, 1, -1, 0) $$
	$$L_{\mu}=\begin{pmatrix}\nu_{\mu}\\
		\mu_{L}\end{pmatrix} \sim (1, 2, -\frac{1}{2}, 1), \; \mu_R \sim (1, 1, -1, 1) $$
	$$L_{\tau}=\begin{pmatrix}\nu_{\tau}\\
		\tau_{L}\end{pmatrix} \sim (1, 2, -\frac{1}{2}, -1), \;  \tau_R \sim (1, 1, -1, -1)$$  \\
	%
	Three right handed neutrinos $N_e, N_{\mu}, N_{\tau}$ with $L_{\mu}-L_{\tau}$ charges $0, 1, -1$ respectively can be introduced along with singlet scalars $\Phi_1, \Phi_2$ with corresponding $L_{\mu}-L_{\tau}$ charges $1, -1$ respectively to take care of spontaneous gauge symmetry breaking and type I seesaw origin of light neutrino masses (see \cite{Patra:2016shz} and references therein for details). Denoting the vacuum expectation values (VEV) of singlets $\Phi_{1,2}$ as $v_{1,2}$, the new gauge boson mass can be found to be $M_{Z'}=g_{\mu \tau} \sqrt{(v^2_1+v^2_2)}$ with $g_{\mu \tau}$ being the $L_{\mu}-L_{\tau}$ gauge coupling. Clearly the model predicts diagonal charged lepton mass matrix $M_\ell$ and diagonal Dirac Yukawa of light neutrinos. Thus, the non-trivial neutrino mixing will arise from the structure of right handed neutrino mass matrix $M_R$ only which is generated by the chosen scalar singlet fields.

	For dark matter sector, we introduce two additional vector like fermions $\psi_{A, B}$ and two additional singlet scalar $\eta$ and $\xi$. The $L_{\mu}-L_{\tau}$ gauge couplings of $\psi_A, \psi_B, \eta, \xi$ are taken to be $0, g_B, 0 ~{\rm and} ~0 $ respectively. While $\eta$ gives rise to non-thermal contribution to $\psi_A$ abundance via late decay, the other scalar $\xi$ is responsible for mediating $\psi_A$ annihilation into $\psi_B$. The relevant Lagrangian can be written as follows.
	\begin{align}
		\mathcal{L} & \supseteq \overline{\psi_A} i \gamma^\mu \partial_\mu \psi_A - m_A \overline{\psi_A} \psi_A +\overline{\psi_B} i \gamma^\mu D_\mu \psi_B - m_B \overline{\psi_B} \psi_B \nonumber \\&-y_A \eta \overline{\psi_A} \psi_A -y_B \eta \overline{\psi_B} \psi_B -y_1 \xi \overline{\psi_A} \psi_A -y_2 \xi \overline{\psi_B} \psi_B+{\rm h.c.}
	\end{align}
	Here $D_\mu \psi_B = (\partial_{\mu}-ig_B Z'_{\mu}) \psi_B$. Here $g_B = n_B g_{\mu \tau}$ with $n_B$ being gauge charge of vector like fermion $\psi_B$. Since $n_B$ can be chosen independently, we keep $g_B$ as a free parameter. For simplicity, we take $y_B=0$ whereas $y_A$ is taken to be very small to realise the desired DM phenomenology to be discussed later. Since the scalar singlet $\eta$ is required to be produced in thermal bath leaving a thermal relic followed by late decay into DM (to be discussed below), we also write down its key interactions as
	\begin{align}
	    -\mathcal{L}_{\eta} &\supset \lambda_{\eta} \eta^4 + \lambda_{\eta H} \eta^2 (H^{\dagger}_1 H_1)+\lambda_{\eta \Phi_1} \eta^2 (\Phi^\dagger_1 \Phi_1)\nonumber\\&+\lambda_{\eta \Phi_2} \eta^2 (\Phi^\dagger_2 \Phi_2)+\lambda_{\eta \xi} \eta^2 \xi^2
	\end{align}
	where $H_1$ denotes the SM Higgs field.
	
	It should be noted that, a kinetic mixing term between $U(1)_Y$ of SM and $U(1)_{L_{\mu}-L_{\tau}}$ of the form $\frac{\epsilon}{2} B^{\alpha \beta} Y_{\alpha \beta}$ can exist in the Lagrangian where $B^{\alpha\beta}=  \partial^{\alpha}X^{\beta}-\partial^{\beta}X^{\alpha}, Y_{\alpha \beta}$ are the field strength tensors of $U(1)_{L_{\mu}-L_{\tau}}, U(1)_Y$ respectively and $\epsilon$ is the mixing parameter. Even if this mixing is considered to be absent in the Lagrangian, it can arise at one loop level with particles charged under both the gauge sectors in the loop. We consider this mixing to be $\epsilon = g_{\mu \tau}/70$. While the phenomenology of muon $(g-2)$, and DM relic in our model is not dependent on this mixing, the XENON1T fit as well as other experimental constraints on the model parameters can crucially depend upon this mixing. We therefore choose it to be small, around the same order as its one loop value. \\
	
	\noindent
	{\bf Anomalous Muon Magnetic Moment}: The magnetic moment of muon is given by
	\begin{equation}\label{anomaly}
		\overrightarrow{\mu_\mu}= g_\mu \left (\frac{q}{2m} \right)
		\overrightarrow{S}\,,
	\end{equation}
	where $g_\mu$ is the gyromagnetic ratio and its value is $2$ for a
	structureless, spin $\frac{1}{2}$ particle of mass $m$ and charge
	$q$. Any radiative correction, which couples the muon spin to the
	virtual fields, contributes to its magnetic moment and is given by
	\begin{equation}
		a_\mu=\frac{1}{2} ( g_\mu - 2)
	\end{equation}
	The anomalous muon magnetic moment has been measured very precisely while it has also been predicted in the SM to a great accuracy.
	In our model, the additional contribution to muon magnetic moment comes from one loop diagram mediated by $Z'$ boson. The contribution is given by \cite{Brodsky:1967sr, Baek:2008nz, Queiroz:2014zfa}
	\begin{equation}
		\Delta a_{\mu} = \frac{\alpha'}{2\pi} \int^1_0 dx \frac{2m^2_{\mu} x^2 (1-x)}{x^2 m^2_{\mu}+(1-x)M^2_{Z'}} \approx \frac{\alpha'}{2\pi} \frac{2m^2_{\mu}}{3M^2_{Z'}}
	\end{equation}
	where $\alpha'=g^2_{\mu\tau}/(4\pi)$. \\
	
\begin{figure}[h!]
\centering	\includegraphics[scale=0.45]{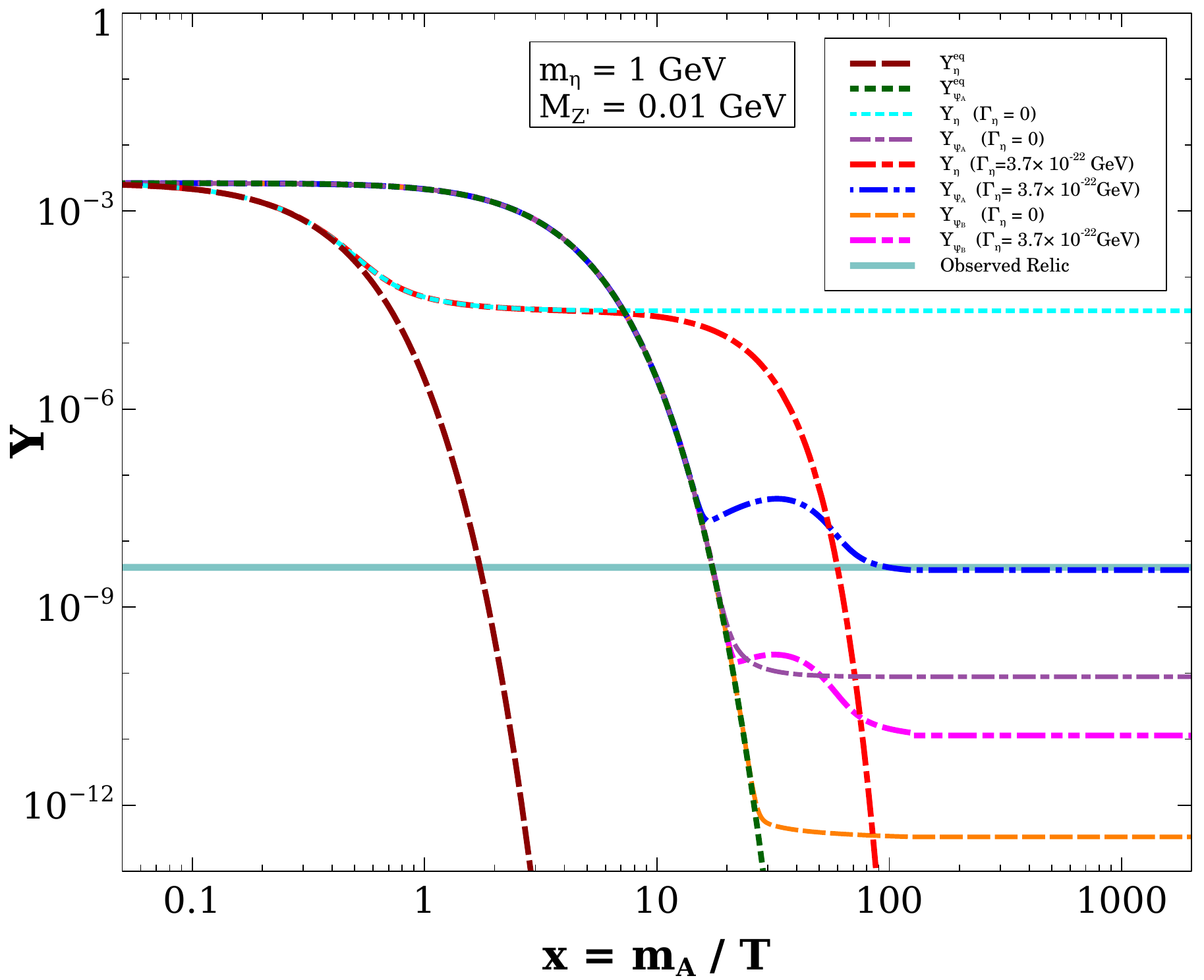}
		\caption{Comoving number densities of DM candidates $\psi_{A,B}$ and long-lived scalar singlet $\eta$. The value of $ \sigma(\eta \eta \rightarrow XX)$ used in the calculation is $1.45\times 10^{-12}$GeV$^{-2}$ which can be obtained for $m_{\eta}=1$ GeV,$m_{X}=0.2$ GeV and the corresponding coupling  $10^{-5}$ in Eq.\ref{etacrs}  }
		\label{figdm}
	\end{figure}

	\noindent
	{\bf Relic Abundance of DM}: 
	Among the dark sector particles, we consider $\psi_A$ to be the dominant DM candidate whose thermal relic is dictated by its annihilation cross section into a pair of $\psi_B$. The scalar singlet $\eta$ is also thermally produced but long lived and it decays at late epochs to $\psi_A$ giving a non-thermal relic contribution. Since $\eta$ coupling to $\psi_A$ is very small from such late decay criteria, it relies upon scalar portal couplings to enter thermal equilibrium with the bath, similar to scalar singlet DM. With these in mind, we can now write down the Boltzmann equations for the DM candidate $\psi_{A}$ along with $\psi_B$ and the scalar singlet $\eta$ whose late decays into DM is crucial to generate correct DM relic. Both the DM and the scalar singlet were in thermal equilibrium in the early universe. Due to large annihilation cross-section $\sigma(\psi_A \psi_A \to \psi_B \psi_B)$, the thermal freeze-out relic of DM is several orders of magnitude smaller than the observed relic density. However the relic can be lifted up to the correct ballpark from late decays of the scalar $\eta$ into DM. The abundance of $\psi_B$ is naturally suppressed due to its large interaction rate with $L_{\mu}-L_{\tau}$ gauge sector. For earlier works on  interplay of $\psi_A \psi_A \to \psi_B \psi_B$ and $\psi_B \psi_B \to {\rm SM \; SM}$ one may refer to \cite{Belanger:2011ww}. 
We define comoving number densities of these particles as $Y_{\psi_A}=n_{\psi_{A}}/s,Y_{\psi_B}=n_{\psi_{B}}/s \; {\rm and} \; Y_{\eta} = n_{\eta}/s$. The relevant coupled Boltzmann equations can then be written as
\begin{widetext}
		\begin{equation}\label{boltzmann}
			\begin{aligned}
				\frac{dY_{\psi_A}}{dx}&=-\frac{s(m_A)}{x^2  H(m_A)} \langle \sigma(\psi_A \psi_A \to \psi_B \psi_B) v\rangle \Big(Y^2_{\psi_A} - \frac{(Y^{\rm eq}_{\psi_A})^2}{(Y^{\rm eq}_{\psi_B})^2}Y_{\psi_B}^2\Big)-\frac{s(m_A)}{x^2  H(m_A)} \langle \sigma(\psi_A \psi_A \to \xi \xi) v \rangle \big(Y^2_{\psi_A} - (Y^{\rm eq}_{\psi_A}\big)^2\big) \nonumber\\&+ \frac{2x}{H(m_A)}\big( \langle \Gamma_{\eta\rightarrow \psi_A \psi_A}\rangle Y_{\eta} \big)\\
				\frac{dY_{\psi_B}}{dx}&=+\frac{s(m_A)}{x^2  H(m_A)} \langle \sigma(\psi_A \psi_A \to \psi_B \psi_B) v \rangle \Big(Y^2_{\psi_A} - \frac{(Y^{\rm eq}_{\psi_A})^2}{(Y^{\rm eq}_{\psi_B})^2}Y_{\psi_B}^2\Big) -\frac{s(m_A)}{x^2  H(m_A)} \langle \sigma(\psi_B \psi_B \to P P) v \rangle \big(Y^2_{\psi_B} - (Y^{\rm eq}_{\psi_B}\big)^2\big)
				\nonumber \\&-\frac{s(m_A)}{x^2  H(m_A)} \langle \sigma(\psi_B \psi_B \to {\rm SM~SM}) v \rangle \big(Y^2_{\psi_B} - (Y^{\rm eq}_{\psi_B}\big)^2\big) \\
				\frac{dY_{\eta}}{dx}&= -\frac{s(m_A)}{x^2  H(m_A)} \langle \sigma(\eta \eta \to XX) v \rangle \big(Y^2_{\eta} - (Y^{\rm eq}_{\eta}\big)^2\big)- \frac{2x}{H(m_A)}\big( \langle \Gamma_{\eta\rightarrow \psi_A \psi_A}\rangle Y_{\eta} \big)
			\end{aligned}
		\end{equation}	
	\end{widetext}
	where,  $x=\frac{m_{A}}{T}$, $s(m_A)=\frac{2\pi^2}{45}g_{*s}m^3_{A}$ and $H(m_A)=1.67 g^{1/2}_*\frac{m^2_A}{M_{Pl}}$. 
We show a benchmark plot in figure \ref{figdm} with $m_\eta$ = 1 GeV, $y_A = 10^{-10}$, $m_A=0.1$ GeV, $m_B = 0.099875$ GeV, $M_{Z'} = 0.01 $~GeV. We consider a large $\sigma(\psi_A \psi_A \to \psi_B \psi_B)$ cross section due to resonance enhancement $2 m_{A} = m_{\xi}$. The reason behind choosing such a large cross section $\sigma(\psi_A \psi_A \to \psi_B \psi_B)$ will become clear when we discuss the XENON1T fit. Here, we have kept $\sigma(\psi_A \psi_A \to \psi_B \psi_B), \sigma(\eta \eta \to X X)$ as free parameters (within unitarity limits) and adjust them to achieve the desired XENON1T fit and DM relic. For example, the relevant couplings can be adjusted to realise such cross sections. All these relevant cross-sections and decay widths are given in the Appendix \ref{appen1}. Note that the decay width $\Gamma_{\eta\rightarrow \psi_A \psi_A}$ is assumed to be very small leading to conversion of $\eta$ into DM during the epoch of the big bang nucleosynthesis (BBN), but well before recombination. In fact, the chosen decay ($\Gamma_{\eta}= 3.7\times 10^{-22}$ GeV) corresponds to a lifetime of approximately $1.7 \times 10^{-3}$ s. This can be still safe from cosmological point of view by forbidding $\eta$ decay into visible sector particles \cite{DEramo:2018khz}. In figure~\ref{figdm}, the dashed brown coloured line shows the equilibrium number density of the singlet scalar $\eta$. This singlet scalar was initially in thermal equilibrium with the SM bath. As its interaction rates falls below the expansion rate, it freezes out leaving a thermal relic, shown by the cyan dotted line. The DM particles $\psi_{A, B}$ were also in thermal equilibrium initially and their equilibrium number densities are shown by the green dashed line (they overlap due to very similar masses). As its interaction rate falls below Hubble rate of expansion, thermal freeze-out occurs leading to abundance of $\psi_{A}$ several orders of magnitude smaller than the observed relic because of its large annihilation cross-section to $\psi_{B}$. This is shown by the purple dot-dashed line. The corresponding thermal freeze-out abundance of $\psi_B$ is also suppressed because of its dominant annihilation into $Z' Z'$ which is shown by the orange dashed line.  The blue dot dashed line depicts the evolution of comoving number density of $\psi_{A}$ after considering the non-thermal contribution from $\eta$ decay. The corresponding depletion of the $\eta$ number density is shown by the red coloured dashed line. Clearly, as the number density of the scalar falls due to its decay, the DM number density gets uplifted. As the comoving number density of $\psi_{A}$ increases, it again starts to annihilate into $\psi_{B} \psi_{B}$, leading to depletion in $\psi_A$ density. Therefore, the final abundance of $\psi_{B}$ which is shown by the dashed pink coloured line, is the result of its production from $\psi_{A} \psi_{A}$ annihilation and its depletion through $\psi_{B}\psi_{B} \rightarrow Z'Z'$. Clearly, even though the final abundance of $\psi_{B}$ is suppressed, the correct relic of $\psi_{A}$ can be obtained by appropriate tuning of the scalar decay width.
	
\begin{figure}[h!]
		\centering
		\includegraphics[scale=0.5]{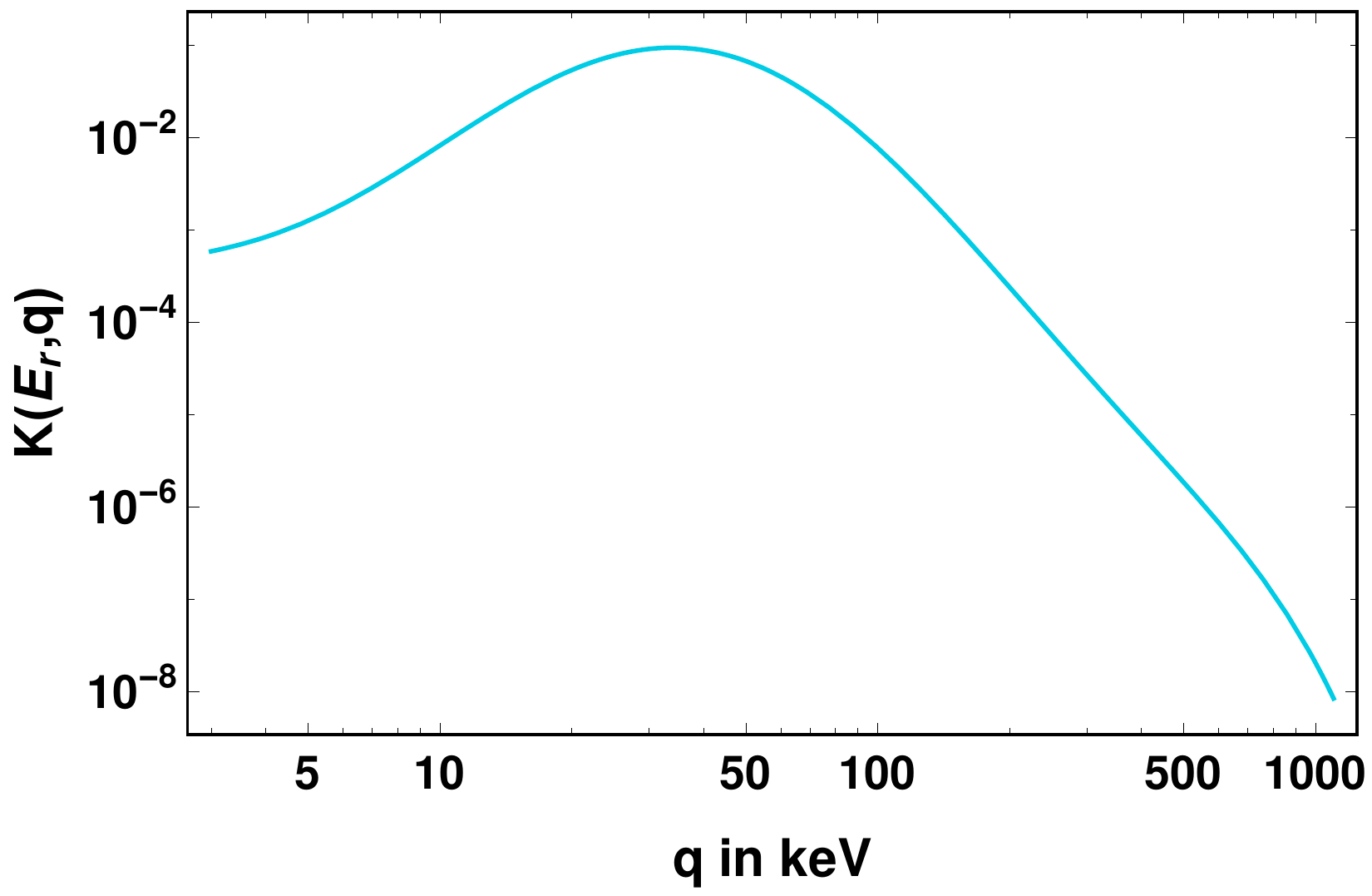}
		\caption{Atomic excitation factor is shown as a function of momentum transferred.}
		\label{aef}
	\end{figure}

\vspace{1cm}	
	\noindent
	{\bf XENON1T Excess}: 
	As mentioned before, we adopt the boosted DM approach in order to explain the XENON1T excess. In this scenario, DM $\psi_A$ annihilates into dark fermion $\psi_B$ giving a significant boost to explain the reported excess in the electron recoil events at XENON1T experiment. For a fixed incoming velocity $v$ of dark fermion, the differential 
	scattering cross-section for the elastic scattering process $\psi_B e \rightarrow 
	\psi_B e$ (with electrons inside the Xenon atom) can be written as
	\begin{equation}
		\frac{d \langle \sigma v \rangle }{d E_r} = \frac{\sigma_e}{2 m_e v } \int_{q-}^{q+} a^2_0 q dq|F(q)|^2 K(E_r,q)\,,
		\label{Event_central}
	\end{equation}	
	where $m_e$ is the electron mass, $\sigma_e$ is the corresponding free electron cross section at fixed momentum transfer 
	$q=1/a_0$ with $a_0 = \frac{1}{\alpha m_e}$ being the Bohr radius, $\alpha = \frac{e^2}{4 \pi}=\frac{1}{137}$ being 
	the fine structure constant, $E_r$ is the recoil energy of electron and $K(E_r, q)$ is the atomic excitation factor. For our calculations, the atomic excitation factor is adopted from \cite{Roberts:2019chv} and we assume the dark fermion form factor to be unity. The dependency of atomic excitation factor on the transferred momentum $q$ is shown in figure~\ref{aef}. Here the dominant contribution comes from the bound states with principal quantum number $n=3$ as their binding energy is around a few keVs.
	
	The free electron scattering cross-section for the process $\psi_B e \rightarrow \psi_B e$ is given by
	\begin{equation}
		\sigma_e = \frac{ g^2_B \epsilon^2 g^2  m^2_e  }{\pi M^4_{Z'}}
		\label{DM-electron-scattering}
	\end{equation}
	$\epsilon$ is the kinetic mixing parameter between $Z$ and $Z'$ gauge bosons, $g$ is the weak gauge coupling and $g_B$ is the gauge coupling between $Z'$ and $\psi_B$ defined earlier. It should be noted here that, for GeV scale dark fermion, $\sigma_e$ 
	is independent of $\psi_B$ mass as the reduced mass of $\psi_B$-electron is almost equal to electron mass. 
	For this elastic scattering the limits of integration are determined from the kinematics and are given by
	
	\begin{equation}
		q_\pm=m_{B} v \pm \sqrt{m^2_{B} v^2 -2m_{B}E_r}\,.
	\end{equation}

	The differential event rate for the scattering of $\psi_B$ with electrons in Xenon atom at XENON1T detector, {\it i.e} $\psi_B e 
	\rightarrow \psi_B e$, can be written as 
	\begin{equation}
		\frac{dR}{dE_r}=n_T \Phi_{\psi_B} \frac{d \langle \sigma v \rangle}{d E_r}
		\label{event_rate}
	\end{equation}
	where $n_T=4\times10^{27}$ $ {\rm Ton}^{-1}$ is the number density of Xenon atoms and $\Phi_{\psi_B}$ is the flux of the boosted $\psi_B$ particle.
	
	Here we consider a scenario with two particles $\psi_A$ and $\psi_B$ where $\psi_A$ is the dominant DM component in the present universe and it's annihilation in DM dense regions like the Galactic center (GC) or the Sun produces boosted $\psi_B$ particles with the boost determined by the mass difference between $\psi_A$ and $\psi_B$. 
	
	If one considers the GC to be the source of boosted dark fermion (via the annihilation of the DM with annihilation cross-section of order $\mathcal{O}(10^{-29}\; {\rm cm}^2)$, then the obtained flux is 
	\begin{equation}
		\Phi^{\rm GC}_{\psi_B}= 1.6 \times 10^5 \; {\rm cm}^{-2} {\rm s}^{-1} \bigg(\frac{\langle\sigma_{ \psi_A\psi_A\rightarrow\psi_B\psi_B}v\rangle}{10^{-29} \; {\rm cm}^2}\bigg)\bigg(\frac{0.1\; {\rm GeV}}{m_A}\bigg)^2
		\label{flux}
	\end{equation}
	
	
	The detected recoil energy spectrum can be obtained by convolving Eq.~\eqref{event_rate} with the energy resolution of the XENON1T 
	detector. Incorporating the detector efficiency $\gamma(E)$, the energy resolution of the detector is given by a Gaussian distribution with an 
	energy dependent width, 
	\begin{equation}
		\zeta(E,E_r)=\frac{1}{\sqrt{2 \pi \sigma^2_{\rm det}}}{\rm Exp}\Big[-\frac{(E-E_r)^2}{2 \sigma^2_{\rm det}}\Big] \times \gamma(E)
	\end{equation}
	where $\gamma(E)$ is reported in figure~2 of \cite{Aprile:2020tmw} and the width $\sigma_{\rm det}$ is given by 
	\begin{equation}
		\sigma_{\rm det}(E)= a \sqrt{E} + b E
	\end{equation} 
	with $a=0.3171$ and $b=0.0037$.
	Thus the final detected recoil energy spectrum is given by
	\begin{equation}
		\frac{dR_{\rm det}}{dE_r}=\frac{n_T \Phi_{\psi_B} \sigma_e a^2_0}{2 m_e v}  \int dE ~~\zeta(E,E_r) \Bigg[\int_{q-}^{q+} dq~~ q K(E_r,q)\Bigg] 	
	\end{equation}
	With the flux mentioned in Eq.\eqref{flux}, the electron scattering cross-section $\sigma_e$ that can explain the electron recoil excess at XENON1T is calculated to be $7.2\times 10^{-11}$GeV$^{-2}$.
	\begin{figure}[h!]
		\centering
		\includegraphics[scale=0.35]{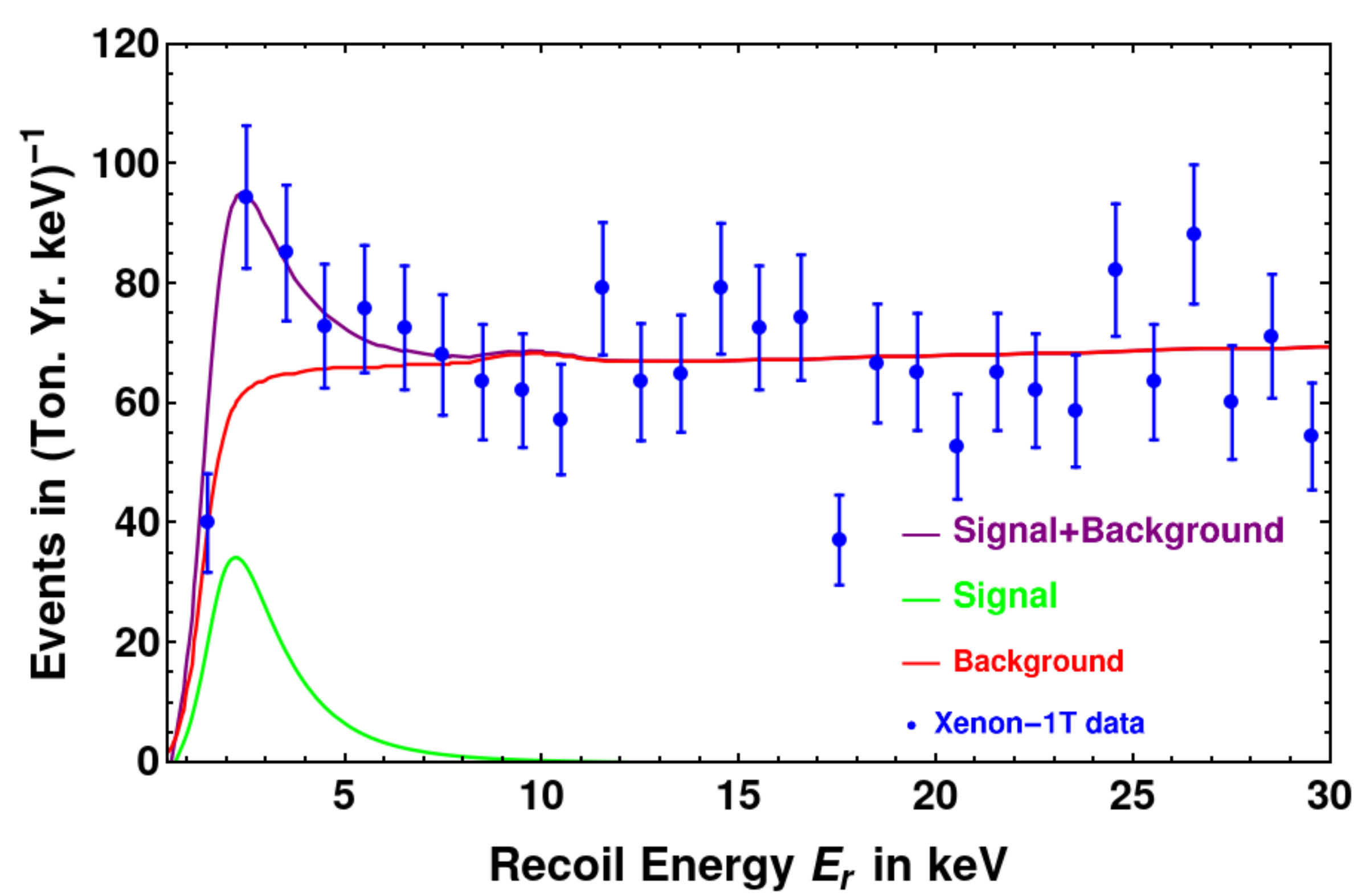}
		\caption{Fit to XENON1T electron recoil excess with the Boosted dark fermion in $L_\mu-L_\tau$ model.}
		\label{xenonfit}
	\end{figure}
	\begin{figure}[h!]
		\centering
		\includegraphics[scale=0.35]{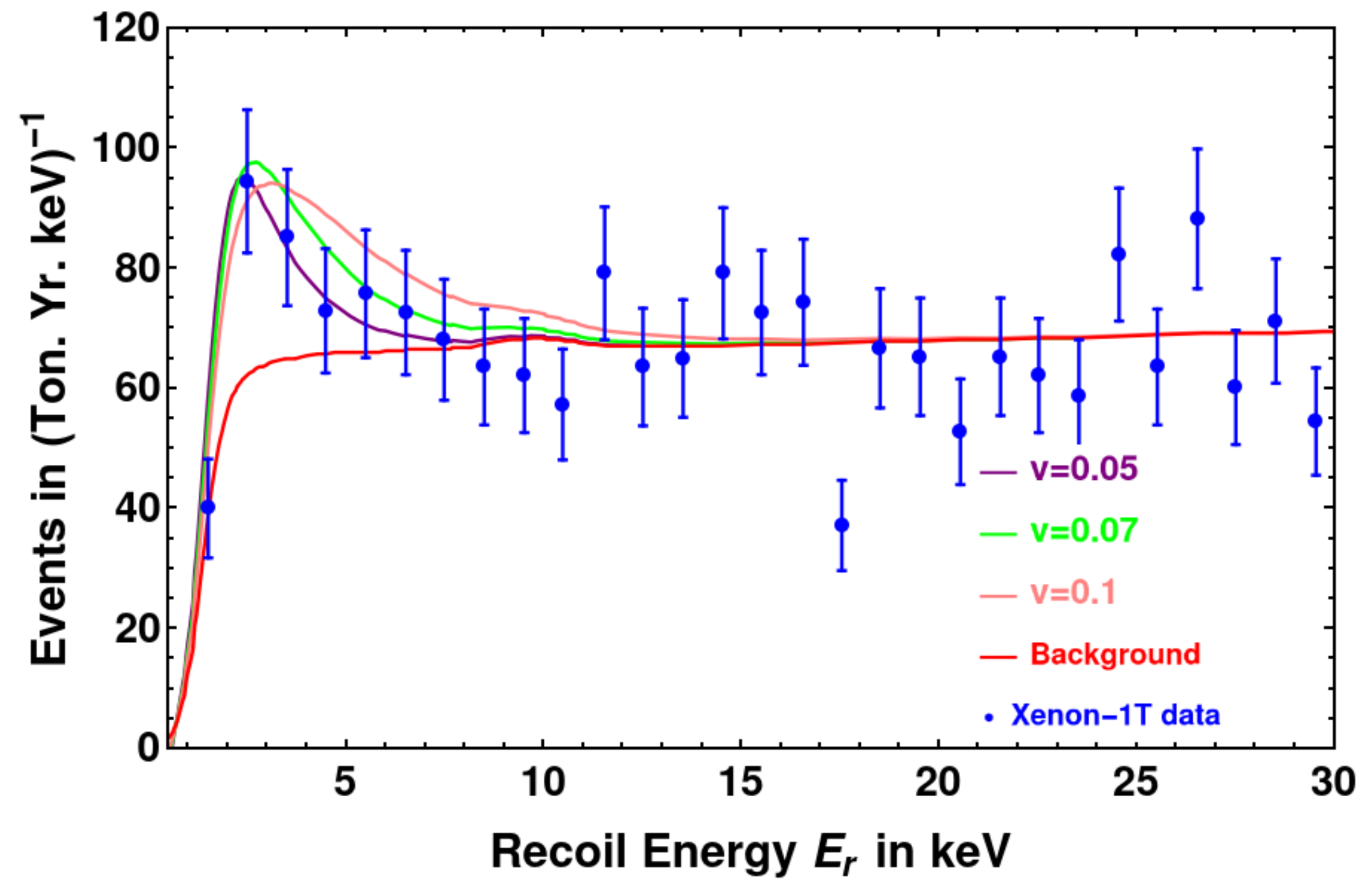}
		\caption{Fit to XENON1T electron recoil excess with different velocity of the boosted dark fermion.}
		\label{xenonfit2}
	\end{figure}
	To obtain the fit to XENON1T data shown in figure \ref{xenonfit} we have used benchmark values $m_B=0.099875$ GeV, $v=0.05$. Such velocities can be obtained by fixing $\Delta m/m_B =1.25\times 10^{-3}$ where $\Delta m = m_A - m_B$ giving rise to the necessary boost factor. In particular we have used  In figure \ref{xenonfit2}, we show the impact of different dark fermion velocities on the fit. Clearly, higher velocities of dark fermion lead to flattening of the fit.
	
	Note that there exists another possibility of getting boosted dark fermion flux from DM annihilation in the Sun. As DM particles can scatter off nuclei inside the Sun and hence gets captured by the Sun, then over certain period of time, DM can get accumulated at the Sun's core. Annihilation of these solar captured DM particles can, in principle, produce lighter boosted particles. In such a scenario the boosted DM flux is no longer dependent on the DM annihilation cross-section but rather it is fully determined by the DM capture rate which is characterised by the DM-nucleon scattering cross-section \cite{Fornal:2020npv}. However, in order to avoid the evaporation bound of a few GeV for DM mass \cite{Griest:1986yu, Gould:1987ju}, we need to choose DM mass in the GeV regime where DM-nucleon scattering rate faces tight constraints from direct search experiments like CRESST-III~\cite{Abdelhameed:2019hmk}. Therefore, one can not get the required enhancement in dark fermion flux from solar captured DM. This justifies the choice of GC as the origin of such DM annihilation into dark fermions. \\

	\noindent
	{\bf Conclusion}:
	\begin{figure}[h!]
		\centering
		\includegraphics[scale=0.5]{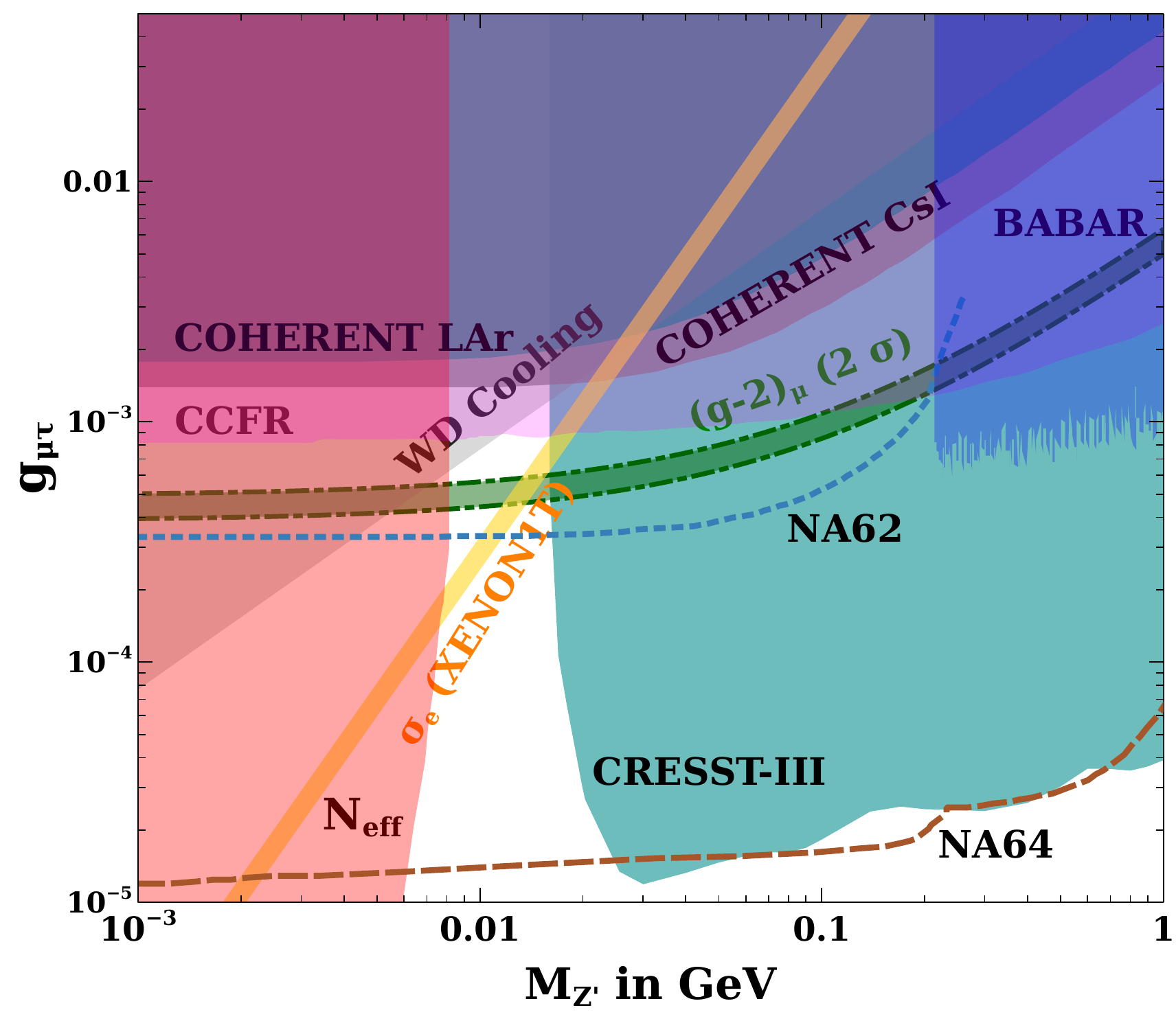}
		\caption{Summary plot showing the final parameter space allowed by all relevant constraints.}
		\label{summary}
	\end{figure}
	
	\begin{figure}[h!]
		\centering
		\includegraphics[scale=0.55]{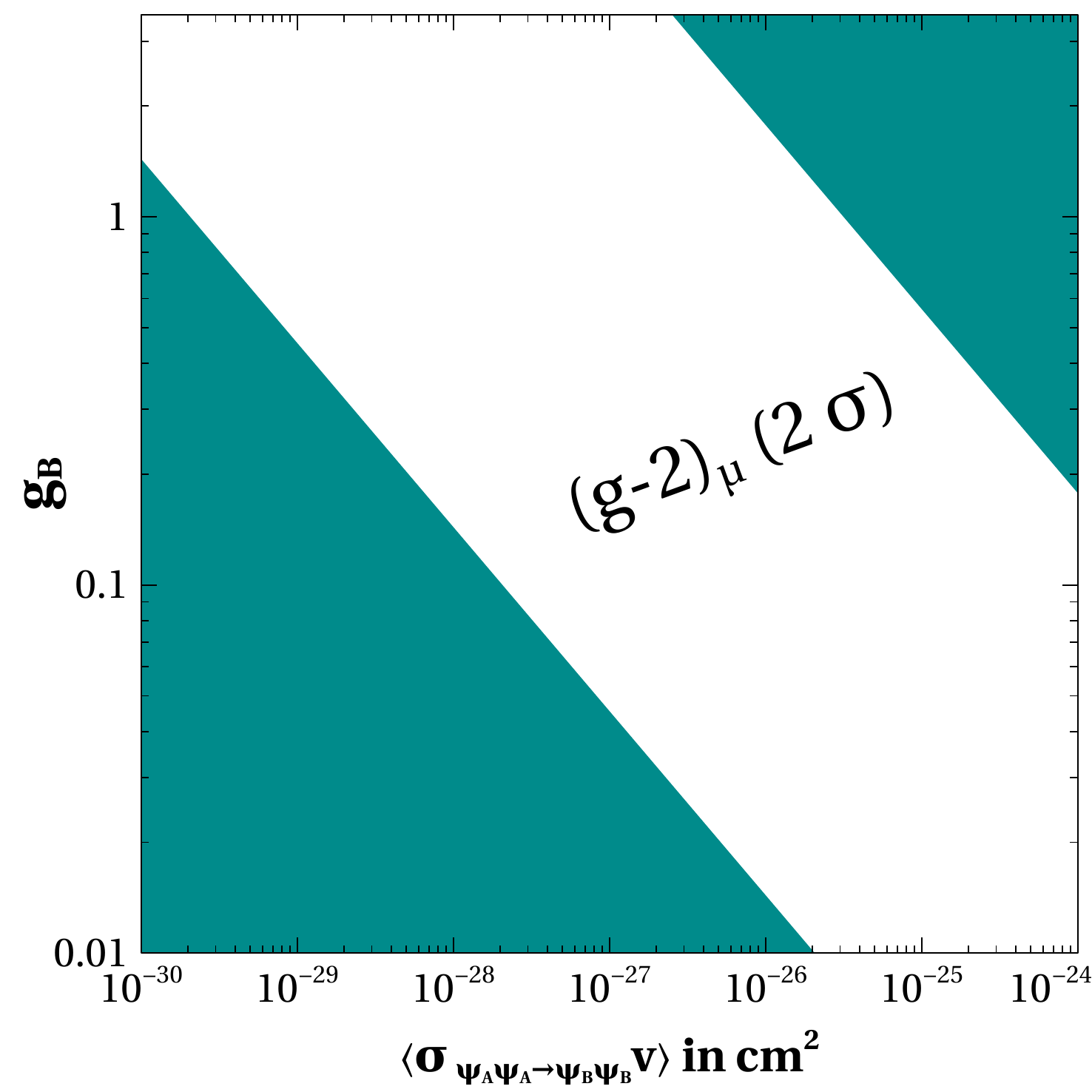}
		\caption{Summary plot showing the final parameter space allowed by all relevant constraints in $g_B$ versus $\sigma_{\psi_A \psi_A \rightarrow \psi_B \psi_B} v$ plane.}
		\label{summary2}
	\end{figure}
	We summarize our result in figure~\ref{summary} in terms of parameter space $g_{\mu\tau}-M_{Z'}$  with kinetic mixing parameter $\epsilon=g_{\mu\tau}/70$. The parameter space satisfying Fermilab's muon $(g-2)$ data is shown by the green coloured band. The orange solid band corresponds to $\sigma_e (\psi_B e \rightarrow \psi_B e) =(6.7-7.7)\times10^{-11} \; {\rm GeV}^{-2}$ required to fit the XENON1T excess for the $\psi_B$ velocity $0.05$ and mass of 0.1 GeV.

	Several experimental constraints are also shown as exclusion bands in the same summary plot of figure \ref{summary}. The strongest constraint ruling out more than half of the $g_{\mu\tau}-M_{Z'}$ plane in figure \ref{summary} comes from low mass DM direct detection experiment CRESST-III~\cite{Abdelhameed:2019hmk}, shown by the green shaded patch. The bound is severe due to the fact that boosted DM $\psi_B$ interacts much more strongly with nucleons via $Z'$ compared to other particles charged under $U(1)_{L_{\mu}-L_{\tau}}$ gauge symmetry. The pink shaded region corresponds to the parameter space excluded by upper bound on cross sections for $\nu N \rightarrow \nu N \mu \bar{\mu}$ measured by CCFR \cite{Altmannshofer:2014pba}. It completely rules out the parameter space satisfying $(g-2)_{\mu}$ beyond $M_{Z'} \gtrsim 0.2$ GeV. The observation of coherent elastic neutrino-nucleus cross section (CE$\nu$NS) in liquid argon (LAr) and cesium-iodide (CsI) performed by the COHERENT Collaboration \cite{Akimov:2017ade, Akimov:2020pdx} also leads to constraint on $g_{\mu \tau}-M_{Z'}$ parameter space. Adopting the analysis of \cite{Cadeddu:2020nbr, Banerjee:2021laz}, we show the constraints from COHERENT LAr and COHERENT CsI by the gray shaded regions. These experiments puts bounds on the CE$\nu$NS cross-section and hence constrains the corresponding coupling and the mediator mass. In the relatively high mass regime of $Z'$, the constraint from the BABAR observations for $4\mu$ final states \cite{TheBABAR:2016rlg}, rules out the parameter space satisfying muon (g-2) beyond $M_{Z'} \gtrsim 0.2$ GeV similar to CCFR as shown by the blue shaded region. The astrophysical bounds from cooling of white dwarf (WD) \cite{Bauer:2018onh, Kamada:2018zxi} excludes the upper left triangular region. Very light $Z'$ is ruled out from cosmological constraints on effective relativistic degrees of freedom \cite{Aghanim:2018eyx, Kamada:2018zxi, Ibe:2019gpv, Escudero:2019gzq}. This arises due to the late decay of such light gauge bosons into SM leptons, after standard neutrino decoupling temperatures thereby enhancing $N_{\rm eff}$. Similar bound also exists for thermal DM masses in this regime which can annihilate into leptons. As shown by the authors of \cite{Sabti:2019mhn}, such constraints from the BBN as well as the cosmic microwave background (CMB) measurements can be satisfied if $M_{\rm DM} \gtrsim \mathcal{O}(1 \; {\rm MeV})$. While in our model we do have some late time annihilations of DM, as seen from figure~\ref{figdm}, such annihilations are from dominant DM candidate $\psi_{A}$ into sub-dominant DM candidate $\psi_{B}$ without involving any visible sector particles in the final state, keeping the scenario safe from BBN bounds. On the other hand, the mass of singlet scalar $\xi$ is kept at $2m_A$ for resonance enhancement of $\psi_A \psi_A \rightarrow \psi_B \psi_B$ cross section and hence heavy enough not to affect BBN.

	We also checked that future experiments at CERN like NA62 \cite{Krnjaic:2019rsv} (blue dashed line of figure~\ref{summary}), and NA64 \cite{Gninenko:2014pea, Gninenko:2018tlp} (brown dashed line in figure~\ref{summary}) is sensitive to our parameter space favoured from DM and muon $(g-2)$ requirements.  Clearly, even after incorporating all existing experimental bounds, there still exists a small parameter space between a few MeV to around 100 MeV consistent with all bounds and the requirement of explaining muon $(g-2)$ and XENON1T excess. Using this final allowed region of parameter space in $g_{\mu\tau}-M_{Z'}$ plane from $(g-2)_{\mu}$ criteria, we show the dark sector parameter space in $g_B$ versus $\sigma_{\psi_A \psi_A \rightarrow \psi_B \psi_B} v$ plane in figure \ref{summary2}. The white colored region is favoured by the latest $(g-2)_{\mu}$ data. The other parameters fixed are $m_A \approx m_B =0.1$ GeV. Clearly, smaller is the annihilation cross section, smaller is the boosted fermion flux and larger is the required coupling $g_B$ to give rise to the XENON1T fit.
	
	Thus, the minimal model we study here is very much constrained from the requirement of $(g-2)_{\mu}$ as well as XENON1T excess from boosted DM. Both visible and dark sector parameters are tightly constrained to a narrow band from these requirements. This is clearly visible from the thin allowed region in figure \ref{summary}. Future data from ongoing and near future experiments can probe the entire parameter space of this minimal and very predictive model. 
	
	\acknowledgements
	DB acknowledges the support from Early Career Research Award from Science and Engineering Research Board (SERB), Department of Science and Technology (DST), Government of India (reference number: ECR/2017/001873). MD acknowledges DST, Government of India for providing the financial assistance for the research under the grant 
	DST/INSPIRE/03/ 2017/000032.
	\appendix
	\begin{widetext}
	\section{Relevant cross-section and decay width}
	\label{appen1}
	\begin{eqnarray}
		\Gamma({\eta\rightarrow \psi_A \psi_A}) &=& \frac{y^2_A}{8\pi}m_{\eta}\Big(1-4\frac{m^2_{A}}{m^2_\eta} \Big)^{3/2}\\
		\sigma({\psi_A \psi_A \rightarrow \psi_B \psi_B})&=&\frac{y^2_1 y^2_2}{32 \pi s }\frac{(s-4m^2_B)^{3/2}(s-4m^2_A)^{1/2}}{(s-m^2_\xi)^2}\\
		\sigma({\rm \psi_B \;\psi_B} \rightarrow Z' Z') &=& \frac{g'^4_B}{192 \pi s (s-4m^2_{B})} \times \Bigg[\frac{24s(4m^4_{B}+2M^4_{Z'}+sm^2_{B})C}{M^4_{Z'}+m^2_{B}s-4M^2_{Z'}m^2_B}\nonumber\\ & -&\frac{24 (8m^2_{B}-4M^2_{Z'}-s^2-(s-2M^2_{Z'})4m^2_{B})}{s-2M^2_{Z'}} {\rm Log}\Big[\frac{2M^2_{Z'}+s(C-1)}{2M^2_{Z'}-s(C+1)}\Big]\Bigg]
	\end{eqnarray}
	where $C=\sqrt{\frac{(s-4M^2_{Z'})(s-4M^2_{\psi_1})}{s^2}}$

	\begin{eqnarray}
\sigma(\psi_{A,B} \psi_{A,B} \rightarrow \xi \xi) &=& \frac{y^4_{1,2}}{32 \pi s (s-4m^2_\psi)}\times \Bigg[\frac{(3m^2_\xi -16 m^2_\xi m^2_\psi+16m^4_psi-2m^2_\psi s) D}{m^4_\xi -2 m^2_\xi m^2_\psi+sm^2_\psi}\nonumber\\&-&\frac{6m^4_\xi-32m^4_\psi+16m^2_\psi s-4m^2_\xi s+16 m^2_\xi m^2_\psi +s^2}{s-2m^2_\xi}Log\Big[\frac{2m^2_\xi+D-s}{2m^2_\xi-D-s}\Big]
\Bigg]
\end{eqnarray}
where $D=\sqrt{(s-4m^2_\xi)(s-4m^2_\psi)}$

\begin{equation}
	\sigma (\eta \eta \to \zeta \zeta)= \frac{\lambda^2_{\eta \zeta}}{64\pi s}\sqrt{\frac{s-4 m^2_{\zeta}}{s-4 m^2_\eta}}
	\label{etacrs}
\end{equation}
where $\zeta$ is any of the singlet scalar lighter than $\eta$.

Thermal averaged cross-section for annihilation of any particle $A$ to $B$  is given by: \cite{Gondolo:1990dk}
\begin{equation}
	\langle\sigma v \rangle_{AA \rightarrow BB} = \frac{x}{2\big[K^2_1(x)+K^2_2(x)\big]}\times \int^{\infty}_{2}  dz \sigma_{(AA\rightarrow BB)} (z^2 m^2_A) (z^2 - 4)z^2 K_1(zx)
\end{equation}
where $z=\sqrt{s}/m_A$ and $x=m_A/T$.

Thermal averaged decay width of $\eta$ decaying to $\psi_{A}$ is given by:
\begin{equation}
	\langle \Gamma(\eta \rightarrow \psi_{A} \psi_{A}) \rangle = \Gamma(\eta \rightarrow \psi_{A} \psi_{A})  \Bigg(\frac{K_1(x)}{K_2(x)}\Bigg)
	\end{equation}

In Eq. A6 and A7, $K_1$ and $K_2$ are the modified Bessel functions of 1st and 2nd kind respectively.
	\end{widetext}

	\bibliographystyle{apsrev}
	\bibstyle{apsrev}
	\bibliography{ref,ref1}

\begin{thebibliography}{111}
\expandafter\ifx\csname natexlab\endcsname\relax\def\natexlab#1{#1}\fi
\expandafter\ifx\csname bibnamefont\endcsname\relax
  \def\bibnamefont#1{#1}\fi
\expandafter\ifx\csname bibfnamefont\endcsname\relax
  \def\bibfnamefont#1{#1}\fi
\expandafter\ifx\csname citenamefont\endcsname\relax
  \def\citenamefont#1{#1}\fi
\expandafter\ifx\csname url\endcsname\relax
  \def\url#1{\texttt{#1}}\fi
\expandafter\ifx\csname urlprefix\endcsname\relax\def\urlprefix{URL }\fi
\providecommand{\bibinfo}[2]{#2}
\providecommand{\eprint}[2][]{\url{#2}}

\bibitem[{\citenamefont{Abi et~al.}(2021)}]{Abi:2021gix}
\bibinfo{author}{\bibfnamefont{B.}~\bibnamefont{Abi}} \bibnamefont{et~al.}
  (\bibinfo{collaboration}{Muon g-2}), \bibinfo{journal}{Phys. Rev. Lett.}
  \textbf{\bibinfo{volume}{126}}, \bibinfo{pages}{141801}
  (\bibinfo{year}{2021}), \eprint{2104.03281}.

\bibitem[{\citenamefont{Aoyama et~al.}(2020)}]{Aoyama:2020ynm}
\bibinfo{author}{\bibfnamefont{T.}~\bibnamefont{Aoyama}} \bibnamefont{et~al.}
  (\bibinfo{year}{2020}), \eprint{2006.04822}.

\bibitem[{\citenamefont{Borsanyi et~al.}(2020)}]{Borsanyi:2020mff}
\bibinfo{author}{\bibfnamefont{S.}~\bibnamefont{Borsanyi}} \bibnamefont{et~al.}
  (\bibinfo{year}{2020}), \eprint{2002.12347}.

\bibitem[{\citenamefont{Zyla et~al.}(2020)}]{Zyla:2020zbs}
\bibinfo{author}{\bibfnamefont{P.~A.} \bibnamefont{Zyla}} \bibnamefont{et~al.}
  (\bibinfo{collaboration}{Particle Data Group}), \bibinfo{journal}{PTEP}
  \textbf{\bibinfo{volume}{2020}}, \bibinfo{pages}{083C01}
  (\bibinfo{year}{2020}).

\bibitem[{\citenamefont{Lindner et~al.}(2018)\citenamefont{Lindner, Platscher,
  and Queiroz}}]{Lindner:2016bgg}
\bibinfo{author}{\bibfnamefont{M.}~\bibnamefont{Lindner}},
  \bibinfo{author}{\bibfnamefont{M.}~\bibnamefont{Platscher}},
  \bibnamefont{and} \bibinfo{author}{\bibfnamefont{F.~S.}
  \bibnamefont{Queiroz}}, \bibinfo{journal}{Phys. Rept.}
  \textbf{\bibinfo{volume}{731}}, \bibinfo{pages}{1} (\bibinfo{year}{2018}),
  \eprint{1610.06587}.

\bibitem[{\citenamefont{Davier et~al.}(2020)\citenamefont{Davier, Hoecker,
  Malaescu, and Zhang}}]{Davier:2019can}
\bibinfo{author}{\bibfnamefont{M.}~\bibnamefont{Davier}},
  \bibinfo{author}{\bibfnamefont{A.}~\bibnamefont{Hoecker}},
  \bibinfo{author}{\bibfnamefont{B.}~\bibnamefont{Malaescu}}, \bibnamefont{and}
  \bibinfo{author}{\bibfnamefont{Z.}~\bibnamefont{Zhang}},
  \bibinfo{journal}{Eur. Phys. J. C} \textbf{\bibinfo{volume}{80}},
  \bibinfo{pages}{241} (\bibinfo{year}{2020}), \bibinfo{note}{[Erratum:
  Eur.Phys.J.C 80, 410 (2020)]}, \eprint{1908.00921}.

\bibitem[{\citenamefont{Davier et~al.}(2017)\citenamefont{Davier, Hoecker,
  Malaescu, and Zhang}}]{Davier:2017zfy}
\bibinfo{author}{\bibfnamefont{M.}~\bibnamefont{Davier}},
  \bibinfo{author}{\bibfnamefont{A.}~\bibnamefont{Hoecker}},
  \bibinfo{author}{\bibfnamefont{B.}~\bibnamefont{Malaescu}}, \bibnamefont{and}
  \bibinfo{author}{\bibfnamefont{Z.}~\bibnamefont{Zhang}},
  \bibinfo{journal}{Eur. Phys. J. C} \textbf{\bibinfo{volume}{77}},
  \bibinfo{pages}{827} (\bibinfo{year}{2017}), \eprint{1706.09436}.

\bibitem[{\citenamefont{Davier et~al.}(2011)\citenamefont{Davier, Hoecker,
  Malaescu, and Zhang}}]{Davier:2010nc}
\bibinfo{author}{\bibfnamefont{M.}~\bibnamefont{Davier}},
  \bibinfo{author}{\bibfnamefont{A.}~\bibnamefont{Hoecker}},
  \bibinfo{author}{\bibfnamefont{B.}~\bibnamefont{Malaescu}}, \bibnamefont{and}
  \bibinfo{author}{\bibfnamefont{Z.}~\bibnamefont{Zhang}},
  \bibinfo{journal}{Eur. Phys. J. C} \textbf{\bibinfo{volume}{71}},
  \bibinfo{pages}{1515} (\bibinfo{year}{2011}), \bibinfo{note}{[Erratum:
  Eur.Phys.J.C 72, 1874 (2012)]}, \eprint{1010.4180}.

\bibitem[{\citenamefont{Arcadi et~al.}(2021{\natexlab{a}})\citenamefont{Arcadi,
  Calibbi, Fedele, and Mescia}}]{Arcadi:2021cwg}
\bibinfo{author}{\bibfnamefont{G.}~\bibnamefont{Arcadi}},
  \bibinfo{author}{\bibfnamefont{L.}~\bibnamefont{Calibbi}},
  \bibinfo{author}{\bibfnamefont{M.}~\bibnamefont{Fedele}}, \bibnamefont{and}
  \bibinfo{author}{\bibfnamefont{F.}~\bibnamefont{Mescia}}
  (\bibinfo{year}{2021}{\natexlab{a}}), \eprint{2104.03228}.

\bibitem[{\citenamefont{Zhu and Liu}(2021)}]{Zhu:2021vlz}
\bibinfo{author}{\bibfnamefont{B.}~\bibnamefont{Zhu}} \bibnamefont{and}
  \bibinfo{author}{\bibfnamefont{X.}~\bibnamefont{Liu}} (\bibinfo{year}{2021}),
  \eprint{2104.03238}.

\bibitem[{\citenamefont{Han et~al.}(2021)\citenamefont{Han, Li, Wang, Wang, and
  Zhang}}]{Han:2021gfu}
\bibinfo{author}{\bibfnamefont{X.-F.} \bibnamefont{Han}},
  \bibinfo{author}{\bibfnamefont{T.}~\bibnamefont{Li}},
  \bibinfo{author}{\bibfnamefont{H.-X.} \bibnamefont{Wang}},
  \bibinfo{author}{\bibfnamefont{L.}~\bibnamefont{Wang}}, \bibnamefont{and}
  \bibinfo{author}{\bibfnamefont{Y.}~\bibnamefont{Zhang}}
  (\bibinfo{year}{2021}), \eprint{2104.03227}.

\bibitem[{\citenamefont{Baum et~al.}(2021)\citenamefont{Baum, Carena, Shah, and
  Wagner}}]{Baum:2021qzx}
\bibinfo{author}{\bibfnamefont{S.}~\bibnamefont{Baum}},
  \bibinfo{author}{\bibfnamefont{M.}~\bibnamefont{Carena}},
  \bibinfo{author}{\bibfnamefont{N.~R.} \bibnamefont{Shah}}, \bibnamefont{and}
  \bibinfo{author}{\bibfnamefont{C.~E.~M.} \bibnamefont{Wagner}}
  (\bibinfo{year}{2021}), \eprint{2104.03302}.

\bibitem[{\citenamefont{Bai and Berger}(2021)}]{Bai:2021bau}
\bibinfo{author}{\bibfnamefont{Y.}~\bibnamefont{Bai}} \bibnamefont{and}
  \bibinfo{author}{\bibfnamefont{J.}~\bibnamefont{Berger}}
  (\bibinfo{year}{2021}), \eprint{2104.03301}.

\bibitem[{\citenamefont{Das et~al.}(2021)\citenamefont{Das, Kumar~Das, and
  Khan}}]{Das:2021zea}
\bibinfo{author}{\bibfnamefont{P.}~\bibnamefont{Das}},
  \bibinfo{author}{\bibfnamefont{M.}~\bibnamefont{Kumar~Das}},
  \bibnamefont{and} \bibinfo{author}{\bibfnamefont{N.}~\bibnamefont{Khan}}
  (\bibinfo{year}{2021}), \eprint{2104.03271}.

\bibitem[{\citenamefont{Lu et~al.}(2021)\citenamefont{Lu, Ramos, and
  Sming~Tsai}}]{Lu:2021vcp}
\bibinfo{author}{\bibfnamefont{C.-T.} \bibnamefont{Lu}},
  \bibinfo{author}{\bibfnamefont{R.}~\bibnamefont{Ramos}}, \bibnamefont{and}
  \bibinfo{author}{\bibfnamefont{Y.-L.} \bibnamefont{Sming~Tsai}}
  (\bibinfo{year}{2021}), \eprint{2104.04503}.

\bibitem[{\citenamefont{Ge et~al.}(2021)\citenamefont{Ge, Ma, and
  Pasquini}}]{Ge:2021cjz}
\bibinfo{author}{\bibfnamefont{S.-F.} \bibnamefont{Ge}},
  \bibinfo{author}{\bibfnamefont{X.-D.} \bibnamefont{Ma}}, \bibnamefont{and}
  \bibinfo{author}{\bibfnamefont{P.}~\bibnamefont{Pasquini}}
  (\bibinfo{year}{2021}), \eprint{2104.03276}.

\bibitem[{\citenamefont{Brdar et~al.}(2021)\citenamefont{Brdar, Jana, Kubo, and
  Lindner}}]{Brdar:2021pla}
\bibinfo{author}{\bibfnamefont{V.}~\bibnamefont{Brdar}},
  \bibinfo{author}{\bibfnamefont{S.}~\bibnamefont{Jana}},
  \bibinfo{author}{\bibfnamefont{J.}~\bibnamefont{Kubo}}, \bibnamefont{and}
  \bibinfo{author}{\bibfnamefont{M.}~\bibnamefont{Lindner}}
  (\bibinfo{year}{2021}), \eprint{2104.03282}.

\bibitem[{\citenamefont{Buen-Abad et~al.}(2021)\citenamefont{Buen-Abad, Fan,
  Reece, and Sun}}]{Buen-Abad:2021fwq}
\bibinfo{author}{\bibfnamefont{M.~A.} \bibnamefont{Buen-Abad}},
  \bibinfo{author}{\bibfnamefont{J.}~\bibnamefont{Fan}},
  \bibinfo{author}{\bibfnamefont{M.}~\bibnamefont{Reece}}, \bibnamefont{and}
  \bibinfo{author}{\bibfnamefont{C.}~\bibnamefont{Sun}} (\bibinfo{year}{2021}),
  \eprint{2104.03267}.

\bibitem[{\citenamefont{Zu et~al.}(2021)\citenamefont{Zu, Pan, Feng, Yuan, and
  Fan}}]{Zu:2021odn}
\bibinfo{author}{\bibfnamefont{L.}~\bibnamefont{Zu}},
  \bibinfo{author}{\bibfnamefont{X.}~\bibnamefont{Pan}},
  \bibinfo{author}{\bibfnamefont{L.}~\bibnamefont{Feng}},
  \bibinfo{author}{\bibfnamefont{Q.}~\bibnamefont{Yuan}}, \bibnamefont{and}
  \bibinfo{author}{\bibfnamefont{Y.-Z.} \bibnamefont{Fan}}
  (\bibinfo{year}{2021}), \eprint{2104.03340}.

\bibitem[{\citenamefont{Amaral et~al.}(2021)\citenamefont{Amaral, Cerde\~no,
  Cheek, and Foldenauer}}]{Amaral:2021rzw}
\bibinfo{author}{\bibfnamefont{D.~W.~P.} \bibnamefont{Amaral}},
  \bibinfo{author}{\bibfnamefont{D.~G.} \bibnamefont{Cerde\~no}},
  \bibinfo{author}{\bibfnamefont{A.}~\bibnamefont{Cheek}}, \bibnamefont{and}
  \bibinfo{author}{\bibfnamefont{P.}~\bibnamefont{Foldenauer}}
  (\bibinfo{year}{2021}), \eprint{2104.03297}.

\bibitem[{\citenamefont{Endo et~al.}(2021)\citenamefont{Endo, Hamaguchi,
  Iwamoto, and Kitahara}}]{Endo:2021zal}
\bibinfo{author}{\bibfnamefont{M.}~\bibnamefont{Endo}},
  \bibinfo{author}{\bibfnamefont{K.}~\bibnamefont{Hamaguchi}},
  \bibinfo{author}{\bibfnamefont{S.}~\bibnamefont{Iwamoto}}, \bibnamefont{and}
  \bibinfo{author}{\bibfnamefont{T.}~\bibnamefont{Kitahara}}
  (\bibinfo{year}{2021}), \eprint{2104.03217}.

\bibitem[{\citenamefont{Ahmed et~al.}(2021)\citenamefont{Ahmed, Khan, Li, Li,
  Raza, and Zhang}}]{Ahmed:2021htr}
\bibinfo{author}{\bibfnamefont{W.}~\bibnamefont{Ahmed}},
  \bibinfo{author}{\bibfnamefont{I.}~\bibnamefont{Khan}},
  \bibinfo{author}{\bibfnamefont{J.}~\bibnamefont{Li}},
  \bibinfo{author}{\bibfnamefont{T.}~\bibnamefont{Li}},
  \bibinfo{author}{\bibfnamefont{S.}~\bibnamefont{Raza}}, \bibnamefont{and}
  \bibinfo{author}{\bibfnamefont{W.}~\bibnamefont{Zhang}}
  (\bibinfo{year}{2021}), \eprint{2104.03491}.

\bibitem[{\citenamefont{Abdughani et~al.}(2021)\citenamefont{Abdughani, Fan,
  Feng, Sming~Tsai, Wu, and Yuan}}]{Abdughani:2021pdc}
\bibinfo{author}{\bibfnamefont{M.}~\bibnamefont{Abdughani}},
  \bibinfo{author}{\bibfnamefont{Y.-Z.} \bibnamefont{Fan}},
  \bibinfo{author}{\bibfnamefont{L.}~\bibnamefont{Feng}},
  \bibinfo{author}{\bibfnamefont{Y.-L.} \bibnamefont{Sming~Tsai}},
  \bibinfo{author}{\bibfnamefont{L.}~\bibnamefont{Wu}}, \bibnamefont{and}
  \bibinfo{author}{\bibfnamefont{Q.}~\bibnamefont{Yuan}}
  (\bibinfo{year}{2021}), \eprint{2104.03274}.

\bibitem[{\citenamefont{Van~Beekveld et~al.}(2021)\citenamefont{Van~Beekveld,
  Beenakker, Schutten, and De~Wit}}]{VanBeekveld:2021tgn}
\bibinfo{author}{\bibfnamefont{M.}~\bibnamefont{Van~Beekveld}},
  \bibinfo{author}{\bibfnamefont{W.}~\bibnamefont{Beenakker}},
  \bibinfo{author}{\bibfnamefont{M.}~\bibnamefont{Schutten}}, \bibnamefont{and}
  \bibinfo{author}{\bibfnamefont{J.}~\bibnamefont{De~Wit}}
  (\bibinfo{year}{2021}), \eprint{2104.03245}.

\bibitem[{\citenamefont{Cox et~al.}(2021)\citenamefont{Cox, Han, and
  Yanagida}}]{Cox:2021gqq}
\bibinfo{author}{\bibfnamefont{P.}~\bibnamefont{Cox}},
  \bibinfo{author}{\bibfnamefont{C.}~\bibnamefont{Han}}, \bibnamefont{and}
  \bibinfo{author}{\bibfnamefont{T.~T.} \bibnamefont{Yanagida}}
  (\bibinfo{year}{2021}), \eprint{2104.03290}.

\bibitem[{\citenamefont{Wang et~al.}(2021{\natexlab{a}})\citenamefont{Wang, Wu,
  Xiao, Yang, and Zhang}}]{Wang:2021bcx}
\bibinfo{author}{\bibfnamefont{F.}~\bibnamefont{Wang}},
  \bibinfo{author}{\bibfnamefont{L.}~\bibnamefont{Wu}},
  \bibinfo{author}{\bibfnamefont{Y.}~\bibnamefont{Xiao}},
  \bibinfo{author}{\bibfnamefont{J.~M.} \bibnamefont{Yang}}, \bibnamefont{and}
  \bibinfo{author}{\bibfnamefont{Y.}~\bibnamefont{Zhang}}
  (\bibinfo{year}{2021}{\natexlab{a}}), \eprint{2104.03262}.

\bibitem[{\citenamefont{Gu et~al.}(2021)\citenamefont{Gu, Liu, Su, and
  Wang}}]{Gu:2021mjd}
\bibinfo{author}{\bibfnamefont{Y.}~\bibnamefont{Gu}},
  \bibinfo{author}{\bibfnamefont{N.}~\bibnamefont{Liu}},
  \bibinfo{author}{\bibfnamefont{L.}~\bibnamefont{Su}}, \bibnamefont{and}
  \bibinfo{author}{\bibfnamefont{D.}~\bibnamefont{Wang}}
  (\bibinfo{year}{2021}), \eprint{2104.03239}.

\bibitem[{\citenamefont{Cao et~al.}(2021)\citenamefont{Cao, Lian, Pan, Zhang,
  and Zhu}}]{Cao:2021tuh}
\bibinfo{author}{\bibfnamefont{J.}~\bibnamefont{Cao}},
  \bibinfo{author}{\bibfnamefont{J.}~\bibnamefont{Lian}},
  \bibinfo{author}{\bibfnamefont{Y.}~\bibnamefont{Pan}},
  \bibinfo{author}{\bibfnamefont{D.}~\bibnamefont{Zhang}}, \bibnamefont{and}
  \bibinfo{author}{\bibfnamefont{P.}~\bibnamefont{Zhu}} (\bibinfo{year}{2021}),
  \eprint{2104.03284}.

\bibitem[{\citenamefont{Yin}(2021)}]{Yin:2021mls}
\bibinfo{author}{\bibfnamefont{W.}~\bibnamefont{Yin}} (\bibinfo{year}{2021}),
  \eprint{2104.03259}.

\bibitem[{\citenamefont{Han}(2021)}]{Han:2021ify}
\bibinfo{author}{\bibfnamefont{C.}~\bibnamefont{Han}} (\bibinfo{year}{2021}),
  \eprint{2104.03292}.

\bibitem[{\citenamefont{Aboubrahim et~al.}(2021)\citenamefont{Aboubrahim,
  Klasen, and Nath}}]{Aboubrahim:2021rwz}
\bibinfo{author}{\bibfnamefont{A.}~\bibnamefont{Aboubrahim}},
  \bibinfo{author}{\bibfnamefont{M.}~\bibnamefont{Klasen}}, \bibnamefont{and}
  \bibinfo{author}{\bibfnamefont{P.}~\bibnamefont{Nath}}
  (\bibinfo{year}{2021}), \eprint{2104.03839}.

\bibitem[{\citenamefont{Yang et~al.}(2021)\citenamefont{Yang, Zhang, Liu, Dong,
  and Feng}}]{Yang:2021duj}
\bibinfo{author}{\bibfnamefont{J.-L.} \bibnamefont{Yang}},
  \bibinfo{author}{\bibfnamefont{H.-B.} \bibnamefont{Zhang}},
  \bibinfo{author}{\bibfnamefont{C.-X.} \bibnamefont{Liu}},
  \bibinfo{author}{\bibfnamefont{X.-X.} \bibnamefont{Dong}}, \bibnamefont{and}
  \bibinfo{author}{\bibfnamefont{T.-F.} \bibnamefont{Feng}}
  (\bibinfo{year}{2021}), \eprint{2104.03542}.

\bibitem[{\citenamefont{Chakraborti et~al.}(2021)\citenamefont{Chakraborti,
  Roszkowski, and Trojanowski}}]{Chakraborti:2021bmv}
\bibinfo{author}{\bibfnamefont{M.}~\bibnamefont{Chakraborti}},
  \bibinfo{author}{\bibfnamefont{L.}~\bibnamefont{Roszkowski}},
  \bibnamefont{and}
  \bibinfo{author}{\bibfnamefont{S.}~\bibnamefont{Trojanowski}}
  (\bibinfo{year}{2021}), \eprint{2104.04458}.

\bibitem[{\citenamefont{Ferreira et~al.}(2021)\citenamefont{Ferreira,
  Gon\c{c}alves, Joaquim, and Sher}}]{Ferreira:2021gke}
\bibinfo{author}{\bibfnamefont{P.~M.} \bibnamefont{Ferreira}},
  \bibinfo{author}{\bibfnamefont{B.~L.} \bibnamefont{Gon\c{c}alves}},
  \bibinfo{author}{\bibfnamefont{F.~R.} \bibnamefont{Joaquim}},
  \bibnamefont{and} \bibinfo{author}{\bibfnamefont{M.}~\bibnamefont{Sher}}
  (\bibinfo{year}{2021}), \eprint{2104.03367}.

\bibitem[{\citenamefont{Wang et~al.}(2021{\natexlab{b}})\citenamefont{Wang,
  Wang, and Zhang}}]{Wang:2021fkn}
\bibinfo{author}{\bibfnamefont{H.-X.} \bibnamefont{Wang}},
  \bibinfo{author}{\bibfnamefont{L.}~\bibnamefont{Wang}}, \bibnamefont{and}
  \bibinfo{author}{\bibfnamefont{Y.}~\bibnamefont{Zhang}}
  (\bibinfo{year}{2021}{\natexlab{b}}), \eprint{2104.03242}.

\bibitem[{\citenamefont{Li et~al.}(2021)\citenamefont{Li, Pei, and
  Zhang}}]{Li:2021poy}
\bibinfo{author}{\bibfnamefont{T.}~\bibnamefont{Li}},
  \bibinfo{author}{\bibfnamefont{J.}~\bibnamefont{Pei}}, \bibnamefont{and}
  \bibinfo{author}{\bibfnamefont{W.}~\bibnamefont{Zhang}}
  (\bibinfo{year}{2021}), \eprint{2104.03334}.

\bibitem[{\citenamefont{Cadeddu
  et~al.}(2021{\natexlab{a}})\citenamefont{Cadeddu, Cargioli, Dordei, Giunti,
  and Picciau}}]{Cadeddu:2021dqx}
\bibinfo{author}{\bibfnamefont{M.}~\bibnamefont{Cadeddu}},
  \bibinfo{author}{\bibfnamefont{N.}~\bibnamefont{Cargioli}},
  \bibinfo{author}{\bibfnamefont{F.}~\bibnamefont{Dordei}},
  \bibinfo{author}{\bibfnamefont{C.}~\bibnamefont{Giunti}}, \bibnamefont{and}
  \bibinfo{author}{\bibfnamefont{E.}~\bibnamefont{Picciau}}
  (\bibinfo{year}{2021}{\natexlab{a}}), \eprint{2104.03280}.

\bibitem[{\citenamefont{Calibbi et~al.}(2021)\citenamefont{Calibbi,
  L\'opez-Ib\'a\~nez, Melis, and Vives}}]{Calibbi:2021qto}
\bibinfo{author}{\bibfnamefont{L.}~\bibnamefont{Calibbi}},
  \bibinfo{author}{\bibfnamefont{M.~L.} \bibnamefont{L\'opez-Ib\'a\~nez}},
  \bibinfo{author}{\bibfnamefont{A.}~\bibnamefont{Melis}}, \bibnamefont{and}
  \bibinfo{author}{\bibfnamefont{O.}~\bibnamefont{Vives}}
  (\bibinfo{year}{2021}), \eprint{2104.03296}.

\bibitem[{\citenamefont{Chen et~al.}(2021{\natexlab{a}})\citenamefont{Chen,
  Wen, Xu, and Zhang}}]{Chen:2021vzk}
\bibinfo{author}{\bibfnamefont{J.}~\bibnamefont{Chen}},
  \bibinfo{author}{\bibfnamefont{Q.}~\bibnamefont{Wen}},
  \bibinfo{author}{\bibfnamefont{F.}~\bibnamefont{Xu}}, \bibnamefont{and}
  \bibinfo{author}{\bibfnamefont{M.}~\bibnamefont{Zhang}}
  (\bibinfo{year}{2021}{\natexlab{a}}), \eprint{2104.03699}.

\bibitem[{\citenamefont{Escribano et~al.}(2021)\citenamefont{Escribano,
  Terol-Calvo, and Vicente}}]{Escribano:2021css}
\bibinfo{author}{\bibfnamefont{P.}~\bibnamefont{Escribano}},
  \bibinfo{author}{\bibfnamefont{J.}~\bibnamefont{Terol-Calvo}},
  \bibnamefont{and} \bibinfo{author}{\bibfnamefont{A.}~\bibnamefont{Vicente}}
  (\bibinfo{year}{2021}), \eprint{2104.03705}.

\bibitem[{\citenamefont{Chun and Mondal}(2021)}]{Chun:2021dwx}
\bibinfo{author}{\bibfnamefont{E.~J.} \bibnamefont{Chun}} \bibnamefont{and}
  \bibinfo{author}{\bibfnamefont{T.}~\bibnamefont{Mondal}}
  (\bibinfo{year}{2021}), \eprint{2104.03701}.

\bibitem[{\citenamefont{Arcadi et~al.}(2021{\natexlab{b}})\citenamefont{Arcadi,
  De~Jesus, De~Melo, Queiroz, and Villamizar}}]{Arcadi:2021yyr}
\bibinfo{author}{\bibfnamefont{G.}~\bibnamefont{Arcadi}},
  \bibinfo{author}{\bibfnamefont{A.~S.} \bibnamefont{De~Jesus}},
  \bibinfo{author}{\bibfnamefont{T.~B.} \bibnamefont{De~Melo}},
  \bibinfo{author}{\bibfnamefont{F.~S.} \bibnamefont{Queiroz}},
  \bibnamefont{and} \bibinfo{author}{\bibfnamefont{Y.~S.}
  \bibnamefont{Villamizar}} (\bibinfo{year}{2021}{\natexlab{b}}),
  \eprint{2104.04456}.

\bibitem[{\citenamefont{Chen et~al.}(2021{\natexlab{b}})\citenamefont{Chen,
  Chiang, and Nomura}}]{Chen:2021jok}
\bibinfo{author}{\bibfnamefont{C.-H.} \bibnamefont{Chen}},
  \bibinfo{author}{\bibfnamefont{C.-W.} \bibnamefont{Chiang}},
  \bibnamefont{and} \bibinfo{author}{\bibfnamefont{T.}~\bibnamefont{Nomura}}
  (\bibinfo{year}{2021}{\natexlab{b}}), \eprint{2104.03275}.

\bibitem[{\citenamefont{Nomura and Okada}(2021)}]{Nomura:2021oeu}
\bibinfo{author}{\bibfnamefont{T.}~\bibnamefont{Nomura}} \bibnamefont{and}
  \bibinfo{author}{\bibfnamefont{H.}~\bibnamefont{Okada}}
  (\bibinfo{year}{2021}), \eprint{2104.03248}.

\bibitem[{\citenamefont{Athron et~al.}(2021)\citenamefont{Athron, Bal\'azs,
  Jacob, Kotlarski, St\"ockinger, and St\"ockinger-Kim}}]{Athron:2021iuf}
\bibinfo{author}{\bibfnamefont{P.}~\bibnamefont{Athron}},
  \bibinfo{author}{\bibfnamefont{C.}~\bibnamefont{Bal\'azs}},
  \bibinfo{author}{\bibfnamefont{D.~H.} \bibnamefont{Jacob}},
  \bibinfo{author}{\bibfnamefont{W.}~\bibnamefont{Kotlarski}},
  \bibinfo{author}{\bibfnamefont{D.}~\bibnamefont{St\"ockinger}},
  \bibnamefont{and}
  \bibinfo{author}{\bibfnamefont{H.}~\bibnamefont{St\"ockinger-Kim}}
  (\bibinfo{year}{2021}), \eprint{2104.03691}.

\bibitem[{\citenamefont{Aaij et~al.}(2021)}]{Aaij:2021vac}
\bibinfo{author}{\bibfnamefont{R.}~\bibnamefont{Aaij}} \bibnamefont{et~al.}
  (\bibinfo{collaboration}{LHCb}) (\bibinfo{year}{2021}), \eprint{2103.11769}.

\bibitem[{\citenamefont{Aprile et~al.}(2020)}]{Aprile:2020tmw}
\bibinfo{author}{\bibfnamefont{E.}~\bibnamefont{Aprile}} \bibnamefont{et~al.}
  (\bibinfo{collaboration}{XENON}) (\bibinfo{year}{2020}), \eprint{2006.09721}.

\bibitem[{\citenamefont{Takahashi et~al.}(2020)\citenamefont{Takahashi, Yamada,
  and Yin}}]{Takahashi:2020bpq}
\bibinfo{author}{\bibfnamefont{F.}~\bibnamefont{Takahashi}},
  \bibinfo{author}{\bibfnamefont{M.}~\bibnamefont{Yamada}}, \bibnamefont{and}
  \bibinfo{author}{\bibfnamefont{W.}~\bibnamefont{Yin}},
  \bibinfo{journal}{Phys. Rev. Lett.} \textbf{\bibinfo{volume}{125}},
  \bibinfo{pages}{161801} (\bibinfo{year}{2020}), \eprint{2006.10035}.

\bibitem[{\citenamefont{Alonso-Álvarez
  et~al.}(2020)\citenamefont{Alonso-Álvarez, Ertas, Jaeckel, Kahlhoefer, and
  Thormaehlen}}]{Alonso-Alvarez:2020cdv}
\bibinfo{author}{\bibfnamefont{G.}~\bibnamefont{Alonso-Álvarez}},
  \bibinfo{author}{\bibfnamefont{F.}~\bibnamefont{Ertas}},
  \bibinfo{author}{\bibfnamefont{J.}~\bibnamefont{Jaeckel}},
  \bibinfo{author}{\bibfnamefont{F.}~\bibnamefont{Kahlhoefer}},
  \bibnamefont{and} \bibinfo{author}{\bibfnamefont{L.~J.}
  \bibnamefont{Thormaehlen}} (\bibinfo{year}{2020}), \eprint{2006.11243}.

\bibitem[{\citenamefont{Kannike et~al.}(2020)\citenamefont{Kannike, Raidal,
  Veermäe, Strumia, and Teresi}}]{Kannike:2020agf}
\bibinfo{author}{\bibfnamefont{K.}~\bibnamefont{Kannike}},
  \bibinfo{author}{\bibfnamefont{M.}~\bibnamefont{Raidal}},
  \bibinfo{author}{\bibfnamefont{H.}~\bibnamefont{Veermäe}},
  \bibinfo{author}{\bibfnamefont{A.}~\bibnamefont{Strumia}}, \bibnamefont{and}
  \bibinfo{author}{\bibfnamefont{D.}~\bibnamefont{Teresi}}
  (\bibinfo{year}{2020}), \eprint{2006.10735}.

\bibitem[{\citenamefont{Fornal et~al.}(2020)\citenamefont{Fornal, Sandick, Shu,
  Su, and Zhao}}]{Fornal:2020npv}
\bibinfo{author}{\bibfnamefont{B.}~\bibnamefont{Fornal}},
  \bibinfo{author}{\bibfnamefont{P.}~\bibnamefont{Sandick}},
  \bibinfo{author}{\bibfnamefont{J.}~\bibnamefont{Shu}},
  \bibinfo{author}{\bibfnamefont{M.}~\bibnamefont{Su}}, \bibnamefont{and}
  \bibinfo{author}{\bibfnamefont{Y.}~\bibnamefont{Zhao}},
  \bibinfo{journal}{Phys. Rev. Lett.} \textbf{\bibinfo{volume}{125}},
  \bibinfo{pages}{161804} (\bibinfo{year}{2020}), \eprint{2006.11264}.

\bibitem[{\citenamefont{Du et~al.}(2020)\citenamefont{Du, Liang, Liu, Tran, and
  Xue}}]{Du:2020ybt}
\bibinfo{author}{\bibfnamefont{M.}~\bibnamefont{Du}},
  \bibinfo{author}{\bibfnamefont{J.}~\bibnamefont{Liang}},
  \bibinfo{author}{\bibfnamefont{Z.}~\bibnamefont{Liu}},
  \bibinfo{author}{\bibfnamefont{V.~Q.} \bibnamefont{Tran}}, \bibnamefont{and}
  \bibinfo{author}{\bibfnamefont{Y.}~\bibnamefont{Xue}} (\bibinfo{year}{2020}),
  \eprint{2006.11949}.

\bibitem[{\citenamefont{Ko and Tang}(2020)}]{Ko:2020gdg}
\bibinfo{author}{\bibfnamefont{P.}~\bibnamefont{Ko}} \bibnamefont{and}
  \bibinfo{author}{\bibfnamefont{Y.}~\bibnamefont{Tang}}
  (\bibinfo{year}{2020}), \eprint{2006.15822}.

\bibitem[{\citenamefont{Su et~al.}(2020)\citenamefont{Su, Wang, Wu, Yang, and
  Zhu}}]{Su:2020zny}
\bibinfo{author}{\bibfnamefont{L.}~\bibnamefont{Su}},
  \bibinfo{author}{\bibfnamefont{W.}~\bibnamefont{Wang}},
  \bibinfo{author}{\bibfnamefont{L.}~\bibnamefont{Wu}},
  \bibinfo{author}{\bibfnamefont{J.~M.} \bibnamefont{Yang}}, \bibnamefont{and}
  \bibinfo{author}{\bibfnamefont{B.}~\bibnamefont{Zhu}} (\bibinfo{year}{2020}),
  \eprint{2006.11837}.

\bibitem[{\citenamefont{Harigaya et~al.}(2020)\citenamefont{Harigaya, Nakai,
  and Suzuki}}]{Harigaya:2020ckz}
\bibinfo{author}{\bibfnamefont{K.}~\bibnamefont{Harigaya}},
  \bibinfo{author}{\bibfnamefont{Y.}~\bibnamefont{Nakai}}, \bibnamefont{and}
  \bibinfo{author}{\bibfnamefont{M.}~\bibnamefont{Suzuki}}
  (\bibinfo{year}{2020}), \eprint{2006.11938}.

\bibitem[{\citenamefont{Borah et~al.}(2020{\natexlab{a}})\citenamefont{Borah,
  Mahapatra, Nanda, and Sahu}}]{Borah:2020jzi}
\bibinfo{author}{\bibfnamefont{D.}~\bibnamefont{Borah}},
  \bibinfo{author}{\bibfnamefont{S.}~\bibnamefont{Mahapatra}},
  \bibinfo{author}{\bibfnamefont{D.}~\bibnamefont{Nanda}}, \bibnamefont{and}
  \bibinfo{author}{\bibfnamefont{N.}~\bibnamefont{Sahu}}
  (\bibinfo{year}{2020}{\natexlab{a}}), \eprint{2007.10754}.

\bibitem[{\citenamefont{Choudhury et~al.}(2020)\citenamefont{Choudhury,
  Maharana, Sachdeva, and Sahdev}}]{Choudhury:2020xui}
\bibinfo{author}{\bibfnamefont{D.}~\bibnamefont{Choudhury}},
  \bibinfo{author}{\bibfnamefont{S.}~\bibnamefont{Maharana}},
  \bibinfo{author}{\bibfnamefont{D.}~\bibnamefont{Sachdeva}}, \bibnamefont{and}
  \bibinfo{author}{\bibfnamefont{V.}~\bibnamefont{Sahdev}}
  (\bibinfo{year}{2020}), \eprint{2007.08205}.

\bibitem[{\citenamefont{Bramante and Song}(2020)}]{Bramante:2020zos}
\bibinfo{author}{\bibfnamefont{J.}~\bibnamefont{Bramante}} \bibnamefont{and}
  \bibinfo{author}{\bibfnamefont{N.}~\bibnamefont{Song}},
  \bibinfo{journal}{Phys. Rev. Lett.} \textbf{\bibinfo{volume}{125}},
  \bibinfo{pages}{161805} (\bibinfo{year}{2020}), \eprint{2006.14089}.

\bibitem[{\citenamefont{Bell et~al.}(2020)\citenamefont{Bell, Dent, Dutta,
  Ghosh, Kumar, and Newstead}}]{Bell:2020bes}
\bibinfo{author}{\bibfnamefont{N.~F.} \bibnamefont{Bell}},
  \bibinfo{author}{\bibfnamefont{J.~B.} \bibnamefont{Dent}},
  \bibinfo{author}{\bibfnamefont{B.}~\bibnamefont{Dutta}},
  \bibinfo{author}{\bibfnamefont{S.}~\bibnamefont{Ghosh}},
  \bibinfo{author}{\bibfnamefont{J.}~\bibnamefont{Kumar}}, \bibnamefont{and}
  \bibinfo{author}{\bibfnamefont{J.~L.} \bibnamefont{Newstead}},
  \bibinfo{journal}{Phys. Rev. Lett.} \textbf{\bibinfo{volume}{125}},
  \bibinfo{pages}{161803} (\bibinfo{year}{2020}), \eprint{2006.12461}.

\bibitem[{\citenamefont{Borah et~al.}(2020{\natexlab{b}})\citenamefont{Borah,
  Mahapatra, and Sahu}}]{Borah:2020smw}
\bibinfo{author}{\bibfnamefont{D.}~\bibnamefont{Borah}},
  \bibinfo{author}{\bibfnamefont{S.}~\bibnamefont{Mahapatra}},
  \bibnamefont{and} \bibinfo{author}{\bibfnamefont{N.}~\bibnamefont{Sahu}}
  (\bibinfo{year}{2020}{\natexlab{b}}), \eprint{2009.06294}.

\bibitem[{\citenamefont{Aboubrahim et~al.}(2020)\citenamefont{Aboubrahim,
  Klasen, and Nath}}]{Aboubrahim:2020iwb}
\bibinfo{author}{\bibfnamefont{A.}~\bibnamefont{Aboubrahim}},
  \bibinfo{author}{\bibfnamefont{M.}~\bibnamefont{Klasen}}, \bibnamefont{and}
  \bibinfo{author}{\bibfnamefont{P.}~\bibnamefont{Nath}}
  (\bibinfo{year}{2020}), \eprint{2011.08053}.

\bibitem[{\citenamefont{Lee}(2020)}]{Lee:2020wmh}
\bibinfo{author}{\bibfnamefont{H.~M.} \bibnamefont{Lee}}
  (\bibinfo{year}{2020}), \eprint{2006.13183}.

\bibitem[{\citenamefont{Baek et~al.}(2020)\citenamefont{Baek, Kim, and
  Ko}}]{Baek:2020owl}
\bibinfo{author}{\bibfnamefont{S.}~\bibnamefont{Baek}},
  \bibinfo{author}{\bibfnamefont{J.}~\bibnamefont{Kim}}, \bibnamefont{and}
  \bibinfo{author}{\bibfnamefont{P.}~\bibnamefont{Ko}}, \bibinfo{journal}{Phys.
  Lett. B} \textbf{\bibinfo{volume}{810}}, \bibinfo{pages}{135848}
  (\bibinfo{year}{2020}), \eprint{2006.16876}.

\bibitem[{\citenamefont{Shakeri et~al.}(2020)\citenamefont{Shakeri, Hajkarim,
  and Xue}}]{Shakeri_2020}
\bibinfo{author}{\bibfnamefont{S.}~\bibnamefont{Shakeri}},
  \bibinfo{author}{\bibfnamefont{F.}~\bibnamefont{Hajkarim}}, \bibnamefont{and}
  \bibinfo{author}{\bibfnamefont{S.-S.} \bibnamefont{Xue}},
  \bibinfo{journal}{Journal of High Energy Physics}
  \textbf{\bibinfo{volume}{2020}} (\bibinfo{year}{2020}), ISSN
  \bibinfo{issn}{1029-8479},
  \urlprefix\url{http://dx.doi.org/10.1007/JHEP12(2020)194}.

\bibitem[{\citenamefont{Bally et~al.}(2020)\citenamefont{Bally, Jana, and
  Trautner}}]{Bally:2020yid}
\bibinfo{author}{\bibfnamefont{A.}~\bibnamefont{Bally}},
  \bibinfo{author}{\bibfnamefont{S.}~\bibnamefont{Jana}}, \bibnamefont{and}
  \bibinfo{author}{\bibfnamefont{A.}~\bibnamefont{Trautner}},
  \bibinfo{journal}{Phys. Rev. Lett.} \textbf{\bibinfo{volume}{125}},
  \bibinfo{pages}{161802} (\bibinfo{year}{2020}), \eprint{2006.11919}.

\bibitem[{\citenamefont{Delle~Rose et~al.}(2020)\citenamefont{Delle~Rose,
  Hütsi, Marzo, and Marzola}}]{DelleRose:2020pbh}
\bibinfo{author}{\bibfnamefont{L.}~\bibnamefont{Delle~Rose}},
  \bibinfo{author}{\bibfnamefont{G.}~\bibnamefont{Hütsi}},
  \bibinfo{author}{\bibfnamefont{C.}~\bibnamefont{Marzo}}, \bibnamefont{and}
  \bibinfo{author}{\bibfnamefont{L.}~\bibnamefont{Marzola}}
  (\bibinfo{year}{2020}), \eprint{2006.16078}.

\bibitem[{\citenamefont{Ema et~al.}(2020)\citenamefont{Ema, Sala, and
  Sato}}]{Ema:2020fit}
\bibinfo{author}{\bibfnamefont{Y.}~\bibnamefont{Ema}},
  \bibinfo{author}{\bibfnamefont{F.}~\bibnamefont{Sala}}, \bibnamefont{and}
  \bibinfo{author}{\bibfnamefont{R.}~\bibnamefont{Sato}}
  (\bibinfo{year}{2020}), \eprint{2007.09105}.

\bibitem[{\citenamefont{Dutta et~al.}(2021)\citenamefont{Dutta, Mahapatra,
  Borah, and Sahu}}]{Dutta:2021wbn}
\bibinfo{author}{\bibfnamefont{M.}~\bibnamefont{Dutta}},
  \bibinfo{author}{\bibfnamefont{S.}~\bibnamefont{Mahapatra}},
  \bibinfo{author}{\bibfnamefont{D.}~\bibnamefont{Borah}}, \bibnamefont{and}
  \bibinfo{author}{\bibfnamefont{N.}~\bibnamefont{Sahu}}
  (\bibinfo{year}{2021}), \eprint{2101.06472}.

\bibitem[{\citenamefont{Baryakhtar et~al.}(2020)\citenamefont{Baryakhtar,
  Berlin, Liu, and Weiner}}]{Baryakhtar:2020rwy}
\bibinfo{author}{\bibfnamefont{M.}~\bibnamefont{Baryakhtar}},
  \bibinfo{author}{\bibfnamefont{A.}~\bibnamefont{Berlin}},
  \bibinfo{author}{\bibfnamefont{H.}~\bibnamefont{Liu}}, \bibnamefont{and}
  \bibinfo{author}{\bibfnamefont{N.}~\bibnamefont{Weiner}}
  (\bibinfo{year}{2020}), \eprint{2006.13918}.

\bibitem[{\citenamefont{Chao et~al.}(2020)\citenamefont{Chao, Gao, and
  Jin}}]{Chao:2020yro}
\bibinfo{author}{\bibfnamefont{W.}~\bibnamefont{Chao}},
  \bibinfo{author}{\bibfnamefont{Y.}~\bibnamefont{Gao}}, \bibnamefont{and}
  \bibinfo{author}{\bibfnamefont{M.~j.} \bibnamefont{Jin}}
  (\bibinfo{year}{2020}), \eprint{2006.16145}.

\bibitem[{\citenamefont{An and Yang}(2020)}]{An:2020tcg}
\bibinfo{author}{\bibfnamefont{H.}~\bibnamefont{An}} \bibnamefont{and}
  \bibinfo{author}{\bibfnamefont{D.}~\bibnamefont{Yang}}
  (\bibinfo{year}{2020}), \eprint{2006.15672}.

\bibitem[{\citenamefont{He et~al.}(2020{\natexlab{a}})\citenamefont{He, Wang,
  and Zheng}}]{He:2020wjs}
\bibinfo{author}{\bibfnamefont{H.-J.} \bibnamefont{He}},
  \bibinfo{author}{\bibfnamefont{Y.-C.} \bibnamefont{Wang}}, \bibnamefont{and}
  \bibinfo{author}{\bibfnamefont{J.}~\bibnamefont{Zheng}}
  (\bibinfo{year}{2020}{\natexlab{a}}), \eprint{2007.04963}.

\bibitem[{\citenamefont{Kim et~al.}(2020)\citenamefont{Kim, Nomura, and
  Okada}}]{Kim:2020aua}
\bibinfo{author}{\bibfnamefont{J.}~\bibnamefont{Kim}},
  \bibinfo{author}{\bibfnamefont{T.}~\bibnamefont{Nomura}}, \bibnamefont{and}
  \bibinfo{author}{\bibfnamefont{H.}~\bibnamefont{Okada}},
  \bibinfo{journal}{Phys. Lett. B} \textbf{\bibinfo{volume}{811}},
  \bibinfo{pages}{135862} (\bibinfo{year}{2020}), \eprint{2007.09894}.

\bibitem[{\citenamefont{Keung et~al.}(2020)\citenamefont{Keung, Marfatia, and
  Tseng}}]{Keung:2020uew}
\bibinfo{author}{\bibfnamefont{W.-Y.} \bibnamefont{Keung}},
  \bibinfo{author}{\bibfnamefont{D.}~\bibnamefont{Marfatia}}, \bibnamefont{and}
  \bibinfo{author}{\bibfnamefont{P.-Y.} \bibnamefont{Tseng}}
  (\bibinfo{year}{2020}), \eprint{2009.04444}.

\bibitem[{\citenamefont{He et~al.}(2020{\natexlab{b}})\citenamefont{He, Wang,
  and Zheng}}]{He:2020sat}
\bibinfo{author}{\bibfnamefont{H.-J.} \bibnamefont{He}},
  \bibinfo{author}{\bibfnamefont{Y.-C.} \bibnamefont{Wang}}, \bibnamefont{and}
  \bibinfo{author}{\bibfnamefont{J.}~\bibnamefont{Zheng}}
  (\bibinfo{year}{2020}{\natexlab{b}}), \eprint{2012.05891}.

\bibitem[{\citenamefont{Choi et~al.}(2020)\citenamefont{Choi, Lee, and
  Zhu}}]{Choi:2020ysq}
\bibinfo{author}{\bibfnamefont{S.-M.} \bibnamefont{Choi}},
  \bibinfo{author}{\bibfnamefont{H.~M.} \bibnamefont{Lee}}, \bibnamefont{and}
  \bibinfo{author}{\bibfnamefont{B.}~\bibnamefont{Zhu}} (\bibinfo{year}{2020}),
  \eprint{2012.03713}.

\bibitem[{\citenamefont{McKeen et~al.}(2020)\citenamefont{McKeen, Pospelov, and
  Raj}}]{McKeen:2020vpf}
\bibinfo{author}{\bibfnamefont{D.}~\bibnamefont{McKeen}},
  \bibinfo{author}{\bibfnamefont{M.}~\bibnamefont{Pospelov}}, \bibnamefont{and}
  \bibinfo{author}{\bibfnamefont{N.}~\bibnamefont{Raj}} (\bibinfo{year}{2020}),
  \eprint{2006.15140}.

\bibitem[{\citenamefont{Jho et~al.}(2020)\citenamefont{Jho, Park, Park, and
  Tseng}}]{Jho:2020sku}
\bibinfo{author}{\bibfnamefont{Y.}~\bibnamefont{Jho}},
  \bibinfo{author}{\bibfnamefont{J.-C.} \bibnamefont{Park}},
  \bibinfo{author}{\bibfnamefont{S.~C.} \bibnamefont{Park}}, \bibnamefont{and}
  \bibinfo{author}{\bibfnamefont{P.-Y.} \bibnamefont{Tseng}}
  (\bibinfo{year}{2020}), \eprint{2006.13910}.

\bibitem[{\citenamefont{Alhazmi et~al.}(2020)\citenamefont{Alhazmi, Kim, Kong,
  Mohlabeng, Park, and Shin}}]{Alhazmi:2020fju}
\bibinfo{author}{\bibfnamefont{H.}~\bibnamefont{Alhazmi}},
  \bibinfo{author}{\bibfnamefont{D.}~\bibnamefont{Kim}},
  \bibinfo{author}{\bibfnamefont{K.}~\bibnamefont{Kong}},
  \bibinfo{author}{\bibfnamefont{G.}~\bibnamefont{Mohlabeng}},
  \bibinfo{author}{\bibfnamefont{J.-C.} \bibnamefont{Park}}, \bibnamefont{and}
  \bibinfo{author}{\bibfnamefont{S.}~\bibnamefont{Shin}}
  (\bibinfo{year}{2020}), \eprint{2006.16252}.

\bibitem[{\citenamefont{Das and Sen}(2021)}]{Das:2021lcr}
\bibinfo{author}{\bibfnamefont{A.}~\bibnamefont{Das}} \bibnamefont{and}
  \bibinfo{author}{\bibfnamefont{M.}~\bibnamefont{Sen}} (\bibinfo{year}{2021}),
  \eprint{2104.00027}.

\bibitem[{\citenamefont{He et~al.}(1991{\natexlab{a}})\citenamefont{He, Joshi,
  Lew, and Volkas}}]{He:1990pn}
\bibinfo{author}{\bibfnamefont{X.}~\bibnamefont{He}},
  \bibinfo{author}{\bibfnamefont{G.~C.} \bibnamefont{Joshi}},
  \bibinfo{author}{\bibfnamefont{H.}~\bibnamefont{Lew}}, \bibnamefont{and}
  \bibinfo{author}{\bibfnamefont{R.}~\bibnamefont{Volkas}},
  \bibinfo{journal}{Phys. Rev. D} \textbf{\bibinfo{volume}{43}},
  \bibinfo{pages}{22} (\bibinfo{year}{1991}{\natexlab{a}}).

\bibitem[{\citenamefont{He et~al.}(1991{\natexlab{b}})\citenamefont{He, Joshi,
  Lew, and Volkas}}]{He:1991qd}
\bibinfo{author}{\bibfnamefont{X.-G.} \bibnamefont{He}},
  \bibinfo{author}{\bibfnamefont{G.~C.} \bibnamefont{Joshi}},
  \bibinfo{author}{\bibfnamefont{H.}~\bibnamefont{Lew}}, \bibnamefont{and}
  \bibinfo{author}{\bibfnamefont{R.~R.} \bibnamefont{Volkas}},
  \bibinfo{journal}{Phys. Rev. D} \textbf{\bibinfo{volume}{44}},
  \bibinfo{pages}{2118} (\bibinfo{year}{1991}{\natexlab{b}}).

\bibitem[{\citenamefont{Kim et~al.}(2017)\citenamefont{Kim, Park, and
  Shin}}]{Kim:2016zjx}
\bibinfo{author}{\bibfnamefont{D.}~\bibnamefont{Kim}},
  \bibinfo{author}{\bibfnamefont{J.-C.} \bibnamefont{Park}}, \bibnamefont{and}
  \bibinfo{author}{\bibfnamefont{S.}~\bibnamefont{Shin}},
  \bibinfo{journal}{Phys. Rev. Lett.} \textbf{\bibinfo{volume}{119}},
  \bibinfo{pages}{161801} (\bibinfo{year}{2017}), \eprint{1612.06867}.

\bibitem[{\citenamefont{Giudice et~al.}(2018)\citenamefont{Giudice, Kim, Park,
  and Shin}}]{Giudice:2017zke}
\bibinfo{author}{\bibfnamefont{G.~F.} \bibnamefont{Giudice}},
  \bibinfo{author}{\bibfnamefont{D.}~\bibnamefont{Kim}},
  \bibinfo{author}{\bibfnamefont{J.-C.} \bibnamefont{Park}}, \bibnamefont{and}
  \bibinfo{author}{\bibfnamefont{S.}~\bibnamefont{Shin}},
  \bibinfo{journal}{Phys. Lett. B} \textbf{\bibinfo{volume}{780}},
  \bibinfo{pages}{543} (\bibinfo{year}{2018}), \eprint{1712.07126}.

\bibitem[{\citenamefont{Biswas and Shaw}(2019)}]{Biswas:2019twf}
\bibinfo{author}{\bibfnamefont{A.}~\bibnamefont{Biswas}} \bibnamefont{and}
  \bibinfo{author}{\bibfnamefont{A.}~\bibnamefont{Shaw}},
  \bibinfo{journal}{JHEP} \textbf{\bibinfo{volume}{05}}, \bibinfo{pages}{165}
  (\bibinfo{year}{2019}), \eprint{1903.08745}.

\bibitem[{\citenamefont{Patra et~al.}(2017)\citenamefont{Patra, Rao, Sahoo, and
  Sahu}}]{Patra:2016shz}
\bibinfo{author}{\bibfnamefont{S.}~\bibnamefont{Patra}},
  \bibinfo{author}{\bibfnamefont{S.}~\bibnamefont{Rao}},
  \bibinfo{author}{\bibfnamefont{N.}~\bibnamefont{Sahoo}}, \bibnamefont{and}
  \bibinfo{author}{\bibfnamefont{N.}~\bibnamefont{Sahu}},
  \bibinfo{journal}{Nucl. Phys. B} \textbf{\bibinfo{volume}{917}},
  \bibinfo{pages}{317} (\bibinfo{year}{2017}), \eprint{1607.04046}.

\bibitem[{\citenamefont{Brodsky and De~Rafael}(1968)}]{Brodsky:1967sr}
\bibinfo{author}{\bibfnamefont{S.~J.} \bibnamefont{Brodsky}} \bibnamefont{and}
  \bibinfo{author}{\bibfnamefont{E.}~\bibnamefont{De~Rafael}},
  \bibinfo{journal}{Phys. Rev.} \textbf{\bibinfo{volume}{168}},
  \bibinfo{pages}{1620} (\bibinfo{year}{1968}).

\bibitem[{\citenamefont{Baek and Ko}(2009)}]{Baek:2008nz}
\bibinfo{author}{\bibfnamefont{S.}~\bibnamefont{Baek}} \bibnamefont{and}
  \bibinfo{author}{\bibfnamefont{P.}~\bibnamefont{Ko}}, \bibinfo{journal}{JCAP}
  \textbf{\bibinfo{volume}{10}}, \bibinfo{pages}{011} (\bibinfo{year}{2009}),
  \eprint{0811.1646}.

\bibitem[{\citenamefont{Queiroz and Shepherd}(2014)}]{Queiroz:2014zfa}
\bibinfo{author}{\bibfnamefont{F.~S.} \bibnamefont{Queiroz}} \bibnamefont{and}
  \bibinfo{author}{\bibfnamefont{W.}~\bibnamefont{Shepherd}},
  \bibinfo{journal}{Phys. Rev. D} \textbf{\bibinfo{volume}{89}},
  \bibinfo{pages}{095024} (\bibinfo{year}{2014}), \eprint{1403.2309}.

\bibitem[{\citenamefont{Belanger and Park}(2012)}]{Belanger:2011ww}
\bibinfo{author}{\bibfnamefont{G.}~\bibnamefont{Belanger}} \bibnamefont{and}
  \bibinfo{author}{\bibfnamefont{J.-C.} \bibnamefont{Park}},
  \bibinfo{journal}{JCAP} \textbf{\bibinfo{volume}{1203}}, \bibinfo{pages}{038}
  (\bibinfo{year}{2012}), \eprint{1112.4491}.

\bibitem[{\citenamefont{D'Eramo and Profumo}(2018)}]{DEramo:2018khz}
\bibinfo{author}{\bibfnamefont{F.}~\bibnamefont{D'Eramo}} \bibnamefont{and}
  \bibinfo{author}{\bibfnamefont{S.}~\bibnamefont{Profumo}},
  \bibinfo{journal}{Phys. Rev. Lett.} \textbf{\bibinfo{volume}{121}},
  \bibinfo{pages}{071101} (\bibinfo{year}{2018}), \eprint{1806.04745}.

\bibitem[{\citenamefont{Roberts and Flambaum}(2019)}]{Roberts:2019chv}
\bibinfo{author}{\bibfnamefont{B.}~\bibnamefont{Roberts}} \bibnamefont{and}
  \bibinfo{author}{\bibfnamefont{V.}~\bibnamefont{Flambaum}},
  \bibinfo{journal}{Phys. Rev. D} \textbf{\bibinfo{volume}{100}},
  \bibinfo{pages}{063017} (\bibinfo{year}{2019}), \eprint{1904.07127}.

\bibitem[{\citenamefont{Griest and Seckel}(1987)}]{Griest:1986yu}
\bibinfo{author}{\bibfnamefont{K.}~\bibnamefont{Griest}} \bibnamefont{and}
  \bibinfo{author}{\bibfnamefont{D.}~\bibnamefont{Seckel}},
  \bibinfo{journal}{Nucl. Phys. B} \textbf{\bibinfo{volume}{283}},
  \bibinfo{pages}{681} (\bibinfo{year}{1987}), \bibinfo{note}{[Erratum:
  Nucl.Phys.B 296, 1034--1036 (1988)]}.

\bibitem[{\citenamefont{Gould}(1987)}]{Gould:1987ju}
\bibinfo{author}{\bibfnamefont{A.}~\bibnamefont{Gould}},
  \bibinfo{journal}{Astrophys. J.} \textbf{\bibinfo{volume}{321}},
  \bibinfo{pages}{560} (\bibinfo{year}{1987}).

\bibitem[{\citenamefont{Abdelhameed et~al.}(2019)}]{Abdelhameed:2019hmk}
\bibinfo{author}{\bibfnamefont{A.}~\bibnamefont{Abdelhameed}}
  \bibnamefont{et~al.} (\bibinfo{collaboration}{CRESST}),
  \bibinfo{journal}{Phys. Rev. D} \textbf{\bibinfo{volume}{100}},
  \bibinfo{pages}{102002} (\bibinfo{year}{2019}), \eprint{1904.00498}.

\bibitem[{\citenamefont{Altmannshofer et~al.}(2014)\citenamefont{Altmannshofer,
  Gori, Pospelov, and Yavin}}]{Altmannshofer:2014pba}
\bibinfo{author}{\bibfnamefont{W.}~\bibnamefont{Altmannshofer}},
  \bibinfo{author}{\bibfnamefont{S.}~\bibnamefont{Gori}},
  \bibinfo{author}{\bibfnamefont{M.}~\bibnamefont{Pospelov}}, \bibnamefont{and}
  \bibinfo{author}{\bibfnamefont{I.}~\bibnamefont{Yavin}},
  \bibinfo{journal}{Phys. Rev. Lett.} \textbf{\bibinfo{volume}{113}},
  \bibinfo{pages}{091801} (\bibinfo{year}{2014}), \eprint{1406.2332}.

\bibitem[{\citenamefont{Akimov et~al.}(2017)}]{Akimov:2017ade}
\bibinfo{author}{\bibfnamefont{D.}~\bibnamefont{Akimov}} \bibnamefont{et~al.}
  (\bibinfo{collaboration}{COHERENT}), \bibinfo{journal}{Science}
  \textbf{\bibinfo{volume}{357}}, \bibinfo{pages}{1123} (\bibinfo{year}{2017}),
  \eprint{1708.01294}.

\bibitem[{\citenamefont{Akimov et~al.}(2021)}]{Akimov:2020pdx}
\bibinfo{author}{\bibfnamefont{D.}~\bibnamefont{Akimov}} \bibnamefont{et~al.}
  (\bibinfo{collaboration}{COHERENT}), \bibinfo{journal}{Phys. Rev. Lett.}
  \textbf{\bibinfo{volume}{126}}, \bibinfo{pages}{012002}
  (\bibinfo{year}{2021}), \eprint{2003.10630}.

\bibitem[{\citenamefont{Cadeddu
  et~al.}(2021{\natexlab{b}})\citenamefont{Cadeddu, Cargioli, Dordei, Giunti,
  Li, Picciau, and Zhang}}]{Cadeddu:2020nbr}
\bibinfo{author}{\bibfnamefont{M.}~\bibnamefont{Cadeddu}},
  \bibinfo{author}{\bibfnamefont{N.}~\bibnamefont{Cargioli}},
  \bibinfo{author}{\bibfnamefont{F.}~\bibnamefont{Dordei}},
  \bibinfo{author}{\bibfnamefont{C.}~\bibnamefont{Giunti}},
  \bibinfo{author}{\bibfnamefont{Y.~F.} \bibnamefont{Li}},
  \bibinfo{author}{\bibfnamefont{E.}~\bibnamefont{Picciau}}, \bibnamefont{and}
  \bibinfo{author}{\bibfnamefont{Y.~Y.} \bibnamefont{Zhang}},
  \bibinfo{journal}{JHEP} \textbf{\bibinfo{volume}{01}}, \bibinfo{pages}{116}
  (\bibinfo{year}{2021}{\natexlab{b}}), \eprint{2008.05022}.

\bibitem[{\citenamefont{Banerjee et~al.}(2021)\citenamefont{Banerjee, Dutta,
  and Roy}}]{Banerjee:2021laz}
\bibinfo{author}{\bibfnamefont{H.}~\bibnamefont{Banerjee}},
  \bibinfo{author}{\bibfnamefont{B.}~\bibnamefont{Dutta}}, \bibnamefont{and}
  \bibinfo{author}{\bibfnamefont{S.}~\bibnamefont{Roy}} (\bibinfo{year}{2021}),
  \eprint{2103.10196}.

\bibitem[{\citenamefont{Lees et~al.}(2016)}]{TheBABAR:2016rlg}
\bibinfo{author}{\bibfnamefont{J.}~\bibnamefont{Lees}} \bibnamefont{et~al.}
  (\bibinfo{collaboration}{BaBar}), \bibinfo{journal}{Phys. Rev. D}
  \textbf{\bibinfo{volume}{94}}, \bibinfo{pages}{011102}
  (\bibinfo{year}{2016}), \eprint{1606.03501}.

\bibitem[{\citenamefont{Bauer et~al.}(2020)\citenamefont{Bauer, Foldenauer, and
  Jaeckel}}]{Bauer:2018onh}
\bibinfo{author}{\bibfnamefont{M.}~\bibnamefont{Bauer}},
  \bibinfo{author}{\bibfnamefont{P.}~\bibnamefont{Foldenauer}},
  \bibnamefont{and} \bibinfo{author}{\bibfnamefont{J.}~\bibnamefont{Jaeckel}},
  \bibinfo{journal}{JHEP} \textbf{\bibinfo{volume}{18}}, \bibinfo{pages}{094}
  (\bibinfo{year}{2020}), \eprint{1803.05466}.

\bibitem[{\citenamefont{Kamada et~al.}(2018)\citenamefont{Kamada, Kaneta,
  Yanagi, and Yu}}]{Kamada:2018zxi}
\bibinfo{author}{\bibfnamefont{A.}~\bibnamefont{Kamada}},
  \bibinfo{author}{\bibfnamefont{K.}~\bibnamefont{Kaneta}},
  \bibinfo{author}{\bibfnamefont{K.}~\bibnamefont{Yanagi}}, \bibnamefont{and}
  \bibinfo{author}{\bibfnamefont{H.-B.} \bibnamefont{Yu}},
  \bibinfo{journal}{JHEP} \textbf{\bibinfo{volume}{06}}, \bibinfo{pages}{117}
  (\bibinfo{year}{2018}), \eprint{1805.00651}.

\bibitem[{\citenamefont{Aghanim et~al.}(2018)}]{Aghanim:2018eyx}
\bibinfo{author}{\bibfnamefont{N.}~\bibnamefont{Aghanim}} \bibnamefont{et~al.}
  (\bibinfo{collaboration}{Planck}) (\bibinfo{year}{2018}),
  \eprint{1807.06209}.

\bibitem[{\citenamefont{Ibe et~al.}(2020)\citenamefont{Ibe, Kobayashi,
  Nakayama, and Shirai}}]{Ibe:2019gpv}
\bibinfo{author}{\bibfnamefont{M.}~\bibnamefont{Ibe}},
  \bibinfo{author}{\bibfnamefont{S.}~\bibnamefont{Kobayashi}},
  \bibinfo{author}{\bibfnamefont{Y.}~\bibnamefont{Nakayama}}, \bibnamefont{and}
  \bibinfo{author}{\bibfnamefont{S.}~\bibnamefont{Shirai}},
  \bibinfo{journal}{JHEP} \textbf{\bibinfo{volume}{04}}, \bibinfo{pages}{009}
  (\bibinfo{year}{2020}), \eprint{1912.12152}.

\bibitem[{\citenamefont{Escudero et~al.}(2019)\citenamefont{Escudero, Hooper,
  Krnjaic, and Pierre}}]{Escudero:2019gzq}
\bibinfo{author}{\bibfnamefont{M.}~\bibnamefont{Escudero}},
  \bibinfo{author}{\bibfnamefont{D.}~\bibnamefont{Hooper}},
  \bibinfo{author}{\bibfnamefont{G.}~\bibnamefont{Krnjaic}}, \bibnamefont{and}
  \bibinfo{author}{\bibfnamefont{M.}~\bibnamefont{Pierre}},
  \bibinfo{journal}{JHEP} \textbf{\bibinfo{volume}{03}}, \bibinfo{pages}{071}
  (\bibinfo{year}{2019}), \eprint{1901.02010}.

\bibitem[{\citenamefont{Sabti et~al.}(2020)\citenamefont{Sabti, Alvey,
  Escudero, Fairbairn, and Blas}}]{Sabti:2019mhn}
\bibinfo{author}{\bibfnamefont{N.}~\bibnamefont{Sabti}},
  \bibinfo{author}{\bibfnamefont{J.}~\bibnamefont{Alvey}},
  \bibinfo{author}{\bibfnamefont{M.}~\bibnamefont{Escudero}},
  \bibinfo{author}{\bibfnamefont{M.}~\bibnamefont{Fairbairn}},
  \bibnamefont{and} \bibinfo{author}{\bibfnamefont{D.}~\bibnamefont{Blas}},
  \bibinfo{journal}{JCAP} \textbf{\bibinfo{volume}{01}}, \bibinfo{pages}{004}
  (\bibinfo{year}{2020}), \eprint{1910.01649}.

\bibitem[{\citenamefont{Krnjaic et~al.}(2020)\citenamefont{Krnjaic,
  Marques-Tavares, Redigolo, and Tobioka}}]{Krnjaic:2019rsv}
\bibinfo{author}{\bibfnamefont{G.}~\bibnamefont{Krnjaic}},
  \bibinfo{author}{\bibfnamefont{G.}~\bibnamefont{Marques-Tavares}},
  \bibinfo{author}{\bibfnamefont{D.}~\bibnamefont{Redigolo}}, \bibnamefont{and}
  \bibinfo{author}{\bibfnamefont{K.}~\bibnamefont{Tobioka}},
  \bibinfo{journal}{Phys. Rev. Lett.} \textbf{\bibinfo{volume}{124}},
  \bibinfo{pages}{041802} (\bibinfo{year}{2020}), \eprint{1902.07715}.

\bibitem[{\citenamefont{Gninenko et~al.}(2015)\citenamefont{Gninenko,
  Krasnikov, and Matveev}}]{Gninenko:2014pea}
\bibinfo{author}{\bibfnamefont{S.}~\bibnamefont{Gninenko}},
  \bibinfo{author}{\bibfnamefont{N.}~\bibnamefont{Krasnikov}},
  \bibnamefont{and} \bibinfo{author}{\bibfnamefont{V.}~\bibnamefont{Matveev}},
  \bibinfo{journal}{Phys. Rev. D} \textbf{\bibinfo{volume}{91}},
  \bibinfo{pages}{095015} (\bibinfo{year}{2015}), \eprint{1412.1400}.

\bibitem[{\citenamefont{Gninenko and Krasnikov}(2018)}]{Gninenko:2018tlp}
\bibinfo{author}{\bibfnamefont{S.}~\bibnamefont{Gninenko}} \bibnamefont{and}
  \bibinfo{author}{\bibfnamefont{N.}~\bibnamefont{Krasnikov}},
  \bibinfo{journal}{Phys. Lett. B} \textbf{\bibinfo{volume}{783}},
  \bibinfo{pages}{24} (\bibinfo{year}{2018}), \eprint{1801.10448}.

\bibitem[{\citenamefont{Gondolo and Gelmini}(1991)}]{Gondolo:1990dk}
\bibinfo{author}{\bibfnamefont{P.}~\bibnamefont{Gondolo}} \bibnamefont{and}
  \bibinfo{author}{\bibfnamefont{G.}~\bibnamefont{Gelmini}},
  \bibinfo{journal}{Nucl. Phys.} \textbf{\bibinfo{volume}{B360}},
  \bibinfo{pages}{145} (\bibinfo{year}{1991}).

\end{thebibliography}
\end{document}